\documentclass[superscriptaddress,amsmath,amssymb,aps,prx,notitlepage,twocolumn,floatfix]{revtex4-2}
\usepackage[utf8]{inputenc}
\usepackage{listings}
\usepackage{footmisc}
\usepackage{enumerate}
\usepackage{latexsym}
\usepackage{braket}
\usepackage{graphicx}
\usepackage[colorlinks=true, allcolors=blue]{hyperref}
\usepackage{array}

\usepackage{tikz} 
\usepackage{comment} 
\usetikzlibrary{decorations.pathmorphing}
\usetikzlibrary{arrows,decorations.markings}
\tikzset{snake it/.style={decorate, decoration=snake}}
\usetikzlibrary {arrows.meta}

\newcommand{\ti}[1]{\textsubscript{#1}}

\newcommand{\f}[1]{\boldsymbol{#1}}
\newcommand{\nk}[1]{\mathrm{#1}}

\newcommand{\Rf}{\boldsymbol{R}}
\newcommand{\kf}{\boldsymbol{k}}
\newcommand{\Kf}{\boldsymbol{K}}
\newcommand{\qf}{\boldsymbol{q}}

\newcommand{\ds}{\downarrow}
\newcommand{\us}{\uparrow}

\newcommand{\iwnk}{(i \omega_n, \boldsymbol{k}) }

\begin{document}

\title{Spin Hall conductivity in the Kane-Mele-Hubbard model at finite temperature}

\author{D. Lessnich}
\email{dominik.lessnich@usherbrooke.ca}
\affiliation{ D{\'e}partement de Physique, Institut quantique, and RQMP Universit{\'e} de Sherbrooke, Sherbrooke, Qu{\'e}bec, Canada  J1K 2R1}
\affiliation{Institute for Theoretical Physics, Goethe University Frankfurt, Max-von-Laue-Strasse 1, 60438 Frankfurt am Main, Germany}

\author{C. Gauvin-Ndiaye}
\affiliation{ D{\'e}partement de Physique, Institut quantique, and RQMP Universit{\'e} de Sherbrooke, Sherbrooke, Qu{\'e}bec, Canada  J1K 2R1}

\author{Roser Valent\'\i}
\email{valenti@itp.uni-frankfurt.de}
\affiliation{Institute for Theoretical Physics, Goethe University Frankfurt, Max-von-Laue-Strasse 1, 60438 Frankfurt am Main, Germany}

\author{A.-M.S.~Tremblay}
\email{andre-marie.tremblay@usherbrooke.ca}
\affiliation{ D{\'e}partement de Physique, Institut quantique, and RQMP Universit{\'e} de Sherbrooke, Sherbrooke, Qu{\'e}bec, Canada  J1K 2R1}

\begin{abstract}
The Kane-Mele model is known to show a quantized spin Hall conductivity at zero temperature. Including Hubbard interactions at each site leads to a quantum phase transition to an 
XY antiferromagnet at sufficiently high interaction strength.
Here, we use the two-particle self-consistent approach (TPSC), which we extend to include spin-orbit coupling, to investigate the Kane-Mele-Hubbard model at finite temperature and half-filling. 
TPSC is a weak to intermediate coupling approach capable of calculating a frequency- and momentum-dependent self-energy from spin and charge fluctuations.
We present results for the spin Hall conductivity and correlation lengths for antiferromagnetic spin fluctuations for different values of temperature, spin-orbit coupling and Hubbard interaction. 
The vertex corrections, which here are analogues of Maki-Thompson contributions, show a strong momentum dependence and give a large contribution in the vicinity of the phase transition at all temperatures. Their inclusion is necessary to observe the quantization of the spin Hall conductivity for the interacting system in the zero temperature limit. At finite temperature, increasing the Hubbard interaction leads to a decrease of the spin Hall conductivity.
This decrease can be explained by band-gap renormalization from scattering of electrons on antiferromagnetic spin fluctuations.

\end{abstract}

\maketitle

\section{Introduction}
\label{sec:Introduction}

The spin Hall effect is a physical phenomenon in which particles experience forces perpendicular to their flow direction, but of opposite directions depending on the spin orientation. 
The spin Hall effect was first proposed in 1971~\cite{Yakonov71}. 
Since its proposal, the spin Hall effect has been realized in a variety of systems, including the semiconductors GaAs and InGaAs where the phenomenon is driven by spin-orbit coupling~\cite{Kato2004,wunderlich2005}, in laser light traversing dielectric junctions~\cite{hosten2008} and in cold atom systems in optical lattices~\cite{Beeler2013,Aidelsburger2013}.

Likewise, the quantum spin Hall effect (QSH) is a spin selective version of the quantum Hall effect and describes
a time reversal invariant electronic state with a bulk electronic band gap which hosts a quantized spin Hall conductivity. 
The state was originally proposed by Kane and Mele~\cite{Kane2005a,Kane2005b} for a single layer of graphene, where the intrinsic spin-orbit coupling (SOC) opens a band gap and causes a band inversion making the bands topological~\cite{Kane2005b}. 
The Kane-Mele model introduced in Ref.~\cite{Kane2005a} resembles graphene and consists of a honeycomb lattice with nearest neighbor hopping and SOC.
The spin Hall conductivity in this case is quantized at zero temperature since it corresponds to a Brillouin zone integral over the Berry curvature of the occupied bands.
However, in graphene the size of the intrinsic SOC was found to be of the order of a few microelectronvolt, so that the QSH would only be observable at unrealistically low temperatures~\cite{sichau2019,Yao2007,Min2006,huertas2006}. 
An observable QSH was first successfully predicted and measured in HgTe quantum wells~\cite{konig2007,bernevig2006}.
The observation in other systems followed~\cite{roth2009,tingxin2015,knez2011,wu2018,Tang2017,Fei2017,Collins2018,Island2019,eck_2022,Bauernfeind2021,schmitt2023,Bampoulis2023}, notably, for example, in recent measurements in the graphene analog germanene~\cite{Bampoulis2023} and in bilayer graphene~\cite{Island2019}.
A large spin Hall effect has been observed as well
in AB-stacked MoTe\ti{2}/WSe\ti{2} moiré bilayers~\cite{tao2023giant}. 
Materials that can generate large spin-polarized currents at room temperature are sought after for spintronic applications~\cite{Luqiao2012,Liu2012,zutic2004,sinova2015}. 
One possible application is a memory device in which the torque generated by a spin current is used to switch the magnetization of an adjacent ferromagnetic layer.~\cite{Luqiao2012,Liu2012} 

A common model to describe the QSH effect is the above-mentioned Kane-Mele model~\cite{Kane2005a}. 
In many of the material realizations, however, electronic correlations may influence the behavior of the systems.
In this article we focus therefore on the spin Hall effect with interactions and at finite temperatures. We consider the Kane-Mele-Hubbard (KMH) model which is a generalization of the original Kane-Mele model where an onsite Hubbard interaction $U$ is added. 
So far, the KMH model has been investigated in Refs.~\onlinecite{Lee2011,griset2012,Hutchinson2021,Hohenadler2011,Hohenadler2012,Rachel2010,Reuther2012,Yu2011,Zheng2011,Wu2012,Laubach2014,Hung2013,Richter2021,mai2023} and reviewed in Refs.~\onlinecite{rev_Hohenadler_2013, Rachel_2018, Meng2014}.
At $T=0$ the SOC-$U$-phase diagram consists of a QSH insulating phase at small $U$ values and a XY antiferromagnet at larger $U$ values. An intermediate spin liquid phase initially suggested in Refs.~\onlinecite{Hohenadler2011,Hohenadler2012,Yu2011,Wu2012} could not be found by large scale QMC simulations~\cite{Sorella2012}. 
Previous numerical studies found the metallic edge states to be gapped out by large enough interactions~\cite{Yamaji2011,Yoshida2016} or to spontaneously break time-reversal symmetry and acquire magnetic order~\cite{Zheng2011}.
For a discussion on the related issue of boundary zeros see Ref.~\cite{wagner2023mott}. 
In addition,
previous numerical studies of the QSH in the Bernevig-Hughes-Zhang-Hubbard model at finite temperatures using dynamical mean-field theory (DMFT) found that interactions decrease the spin Hall conductivity~\cite{Yoshida2012}.

Here, we show that nonlocal correlation effects in the KMH model are strong and that the momentum-dependent vertex corrections resulting from them give a significant contribution to the spin Hall conductivity in the vicinity of the phase transition at all temperatures. 
To our knowledge, vertex corrections to the spin Hall conductivity have not been included in finite-temperature calculations before. 
We observe that the sensitivity to temperature of the spin-Hall conductivity is enhanced by the gap renormalization due to interactions. 
We also find the inclusion of vertex corrections to be necessary to observe the quantization of the spin Hall conductivity for the interacting system in the zero temperature limit. 

The above results are obtained using the two-particle self-consistent method (TPSC)~\cite{vilk1994,vilk1996,vilk1997}, which we extended to include SOC, to study the KMH model numerically~\cite{lessnichtbp}. 
TPSC is a weak to intermediate coupling approach to the Hubbard model. 
It gives a good description of long-wavelength spin fluctuations and provides a momentum- and frequency-dependent self-energy from spin and charge fluctuations.
There are a number of extensions of TPSC including combinations with DMFT~\cite{martin2023,zantout2023,simard2023}, TPSC+~\cite{Schaefer2021,gauvinndiaye2023}, disorder~\cite{gauvin2022}, TPSC+GG~\cite{Schaefer2021,simard2022}, multi-site case~\cite{arya2015,Aizawa_Kuroki_Yamada_2015,Ogura_Kuroki_2015,Zantout_Altmeyer_Backes_Valentí_2018}, multi-orbital case~\cite{miyhara2013,zantout2019,Zantout2021}, non-equilibrium~\cite{simard2022,simard2023} and nearest neighbor interaction~\cite{Davoudi2006,Davoudi2007,Davoudi2008}.
The TPSC approach was used previously to study the antiferromagnetic phase transition on the honeycomb lattice and was shown to be in good agreement with QMC calculations~\cite{arya2015}.  
Hence, we expect that this method is appropriate to the study of the Kane-Mele-Hubbard model.

The article is organized as follows. In Sec.~\ref{sec:model}, we introduce the model and discuss previous findings. In Sec.~\ref{sec:method}, we discuss the inclusion of SOC into the TPSC approach. Numerical details about the calculation are given in Sec.~\ref{sec:num_details}. Results for the antiferromagnetic spin correlation lengths, spin Hall conductivity (SHC) with and without vertex corrections and the band gap renormalization are given in Sec.~\ref{sec:results}. Appendix~\ref{sec:trgsigma} discusses the range of applicability of TPSC in the KMH model.

\section{Model}
\label{sec:model}

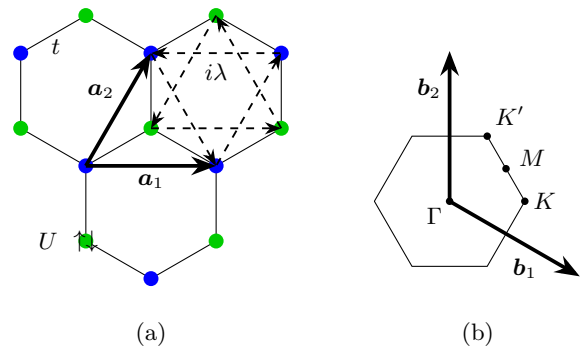
\begin{figure}
\begin{minipage}[b]{.49\linewidth}
    \begin{center}   
    \begin{tikzpicture}[scale=1.]
  \draw [color=black] (0,-1) -- (0.866,-0.5) -- (0.866, 0.5) -- (0,1) -- (-0.866, 0.5) -- (-0.866,-0.5) -- (0,-1);
  \def\a{0.866}
  \draw [color=black] (2*\a -0.866,-0.5) -- (2*\a + 0,-1)-- (2*\a + 0.866,-0.5) -- (2*\a + 0.866, 0.5) -- (2*\a + 0,1) -- (2*\a -0.866, 0.5);
  \def\b{-1.5}
  \draw [color=black] (\a + 0.866, 0.5+\b) -- (\a + 0.866,-0.5+\b) -- (\a + 0,-1+\b) -- (\a -0.866, -0.5+\b)-- (\a -0.866, 0.5+\b);
  
  \node at (0,-1)[color=blue,circle,fill,inner sep=2pt]{};
  \node at (0.866,-0.5)[color=black!20!green,circle,fill,inner sep=2pt]{};
  \node at (0.866, 0.5)[color=blue,circle,fill,inner sep=2pt]{};
  \node at (0,1)[color=black!20!green,circle,fill,inner sep=2pt]{};
  \node at (-0.866, 0.5)[color=blue,circle,fill,inner sep=2pt]{};
  \node at (-0.866,-0.5)[color=black!20!green,circle,fill,inner sep=2pt]{};
  
  \node at (2*\a + 0,-1)[color=blue,circle,fill,inner sep=2pt]{};
  \node at (2*\a + 0.866,-0.5)[color=black!20!green,circle,fill,inner sep=2pt]{};
  \node at (2*\a + 0.866, 0.5)[color=blue,circle,fill,inner sep=2pt]{};
  \node at (2*\a + 0,1)[color=black!20!green,circle,fill,inner sep=2pt]{};
  
  \node at (\a + 0.866,-0.5+\b)[color=black!20!green,circle,fill,inner sep=2pt]{};
  \node at (\a + 0,-1+\b)[color=blue,circle,fill,inner sep=2pt]{};
  \node at (\a -0.866, -0.5+\b)[color=black!20!green,circle,fill,inner sep=2pt]{};

  \draw [color=black,line width=0.5mm,-{Stealth}] (0,-1)-- (2*0.866,-1);
  \draw [color=black,line width=0.5mm,-{Stealth}] (0,-1) -- (0.866,0.5);

  \node at (0.866,-1) [below] {$\f{a}_1$};
  \node at (0.2,0.2) [below] {$\f{a}_2$};
  \node at (-0.4,0.8) [below] {$t$};
  \node at (2*0.866,0.25 )  {$i \lambda$};
  \node at (0,-2)  {$\us \ds $};
  \node at (-0.5,-2)  {$U$};

  \draw [color=black,line width=0.25mm,-{Stealth},dashed] (0.866,-0.5) -- (3*0.866,-0.5);
  \draw [color=black,line width=0.25mm,-{Stealth},dashed] (3*0.866,-0.5) -- (2*0.866,1);
  \draw [color=black,line width=0.25mm,-{Stealth},dashed] (2*0.866,1) -- (0.866,-0.5);

  \draw [color=black,line width=0.25mm,{Stealth}-,dashed] (0.866,0.5) -- (3*0.866,0.5);
  \draw [color=black,line width=0.25mm,{Stealth}-,dashed] (3*0.866,0.5) -- (2*0.866,-1);
  \draw [color=black,line width=0.25mm,{Stealth}-,dashed] (2*0.866,-1) -- (0.866,0.5);
\end{tikzpicture}
\end{center}
(a)
\end{minipage}
\begin{minipage}[b]{.49\linewidth}
\begin{center}
\begin{tikzpicture}[scale=1.]
  \draw [color=black] (-1,0) -- (-0.5, 0.866) -- (0.5, 0.866) -- (1, 0) -- ( 0.5, -0.866) -- (-0.5, -0.866) -- (-1,0);

  \draw [color=black,line width=0.5mm,-{Stealth}] (0,0)-- (0,2);
  \draw [color=black,line width=0.5mm,-{Stealth}] (0,0) -- (2*0.866,-1);
  
  \node at (1.,-0.65 ) [below] {$\f{b}_1$};
  \node at (0.,1.5) [left] {$\f{b}_2$};

  \node at (-0.2,-0.2)  {$\Gamma $};
  \node at (1.,0.) [right] {$K$};
  \node at (0.8,0.9) [above] {$K' $};
  \node at (1.1,0.35) [above] {$M $};

  \node at (0,0)[color=black,circle,fill,inner sep=1pt]{};
  \node at (1.,0)[color=black,circle,fill,inner sep=1pt]{};
  \node at (0.5,0.866 )[color=black,circle,fill,inner sep=1pt]{};
  \node at (0.75,0.5 * 0.866 )[color=black,circle,fill,inner sep=1pt]{};
  
\end{tikzpicture}
\end{center}
(b)
\end{minipage}
\caption{(a) The Kane-Mele-Hubbard model with nearest neighbor hopping $t$, second nearest neigbor spin-orbit coupling $\lambda$ and onsite Hubbard interaction $U$. The basis vectors are $\f{a}_1 = (1,0) a$ and $\f{a}_2 = (1/2,\sqrt{3}/2) a$ with lattice constant $a$. (b) Hexagonal Brillouin zone of the model.}
\label{fig:model}
\end{figure}
The Kane-Mele-Hubbard model on the honeycomb lattice with two sites per unit cell is given by

\begin{align}
    H &= -t \sum_{\braket{i,j}} c^\dagger_i c_j  + i\lambda \sum_{\braket{\braket{i,j}}}  \nu_{ij} c^{\dagger}_i \sigma_z c_j  + U \sum_{i} n_{i \us } n_{i \ds }.
    \label{eq_KM}
\end{align}
Here, the indices $i$ and $j$ run over all lattice sites, and the brackets $\braket{i,j}$ and $\braket{\braket{i,j}}$ denote pairs of nearest and next-nearest neighbors respectively. 
The operator $n_{i\sigma}$ is the number operator for electrons of spin $\sigma$ on site $i$, while $c^{\dagger}_{i }$ ($c_{i } $) is a row (column) vector of creation (annihilation) operators: $c^\dagger_i = (c^{\dagger}_{i \us }, c^{\dagger}_{i \ds })$. 
The next-nearest neighbor hopping parameter $\lambda$ corresponds to the strength of spin-orbit coupling and $\nu_{ij} = \pm 1$  
has a sign that depends on whether going from $i$ to $j$ is clockwise or counterclockwise. 
Finally, $U$ is the on-site Hubbard interaction strength. 
The model and the Brillouin zone are depicted in Fig.~\ref{fig:model}. 
We consider the model at half-filling. The SOC breaks spin rotation symmetry, but time-reversal (TR) symmetry is
preserved and the spin z-component $S^z$ is conserved.

In the noninteracting case the SOC gaps out the Dirac cones at $\kf = \Kf,~\Kf'$ (see Fig.~\ref{fig:model}(b)) with a band gap of $\Delta= 6 \sqrt{3} \lambda$
and at zero temperature the spin Hall conductivity is quantized to a value of $\sigma^{\nk{SH}} = -2 \frac{e^2}{h}$, following the Chern theorem which can be independently applied to spin-up and spin-down subspaces.

At large $U$ values, an XY antiferromagnetic phase is to be expected since the SOC is proportional to $S^z$ making the xy-plane the easy plane for the spins. 
This can be seen in the strong coupling limit of Eq.~\ref{eq_KM} where the Hamiltonian becomes~\cite{Rachel2010}

\begin{equation}
    H_{\infty} = \frac{4 t^2}{U} \sum_{\braket{i,j}} \f{S}_i \cdot \f{S}_j + \frac{4 \lambda^2}{U} \sum_{\braket{\braket{i,j}}} (S_i^z S_j^z - S_i^x S_j^x - S_i^y S_j^y).
\end{equation}

\section{Method}
\label{sec:method}

In the following, we present the multi-band TPSC self-consistency equations including SOC, simplified through conservation of $S^z$. 
The idea of TPSC is to use RPA-like expressions for spin and charge susceptibilities with renormalized two particle vertices, whose values are determined self-consistently using sum rules. 
The self-consistency equations containing SOC are obtained by enforcing time-reversal symmetry. 
So, the system cannot become magnetic in TPSC.
For details regarding the method see Ref.~\onlinecite{lessnichtbp}.

We define the following susceptibilities

\begin{equation}
    \chi_{\alpha \beta}^{ab}(\tau, \Rf_i - \Rf_j ) = \braket{\mathcal{T}_{\tau} O^\alpha_{ia}(\tau ) O^\beta_{jb}(0)} - \braket{ O^\alpha_{ia}} \braket{ O^\beta_{jb}},
    \label{susceptibility}
\end{equation}
where the operators $O^\alpha$ can either be $S^x$, $S^y$, $S^z$ or the number operator $n$ corresponding to the labels $x$, $y$, $z$ or $c$ (charge), $\Rf_i$ is the lattice vector to the unit cell with the index $i$, $a$ and $b$ are site indices labeling the sites in the unit cell, and $\tau$ is the imaginary time.
By Fourier transforming Eq.~\ref{susceptibility} one can go to frequency and momentum space $q = (iq_m, \qf )$, where $q_m=2\pi mT$ is a bosonic Matsubara frequency, with $m$ an integer and $T$ the temperature.
Since $S^z$ is conserved, the longitudinal and the transversal channels decouple. 
From the Bethe-Salpeter equation in the particle-hole channel with a constant irreducible two-particle vertex $\Gamma$, one obtains the following expression for the susceptibilities in the longitudinal particle-hole channel

\begin{equation}
    \chi^l (q) = \left( 1 + \frac{1}{2} \Gamma^l  \chi^l  {}^{(1)} (q) \right)^{-1} \chi^l {}^{(1)} (q) 
    \label{bsel}
\end{equation}
where we defined

\begin{align}
    \chi^l  (q) &= 
\begin{pmatrix}
\chi_{cc}(q) & \chi_{cz}(q)  \\
\chi_{zc}(q) & \chi_{zz}(q)
\end{pmatrix}, \\
\Gamma^l &= 
\begin{pmatrix}
\Gamma_{cc} & 0  \\
0 & - \Gamma_{zz}
\end{pmatrix}.
\end{align}
The superscript $l$ indicates the longitudinal channel and $(1)$ here indicates the corresponding noninteracting expression. 
The two sites in the unit cell are symmetry related so that the vertices are diagonal matrices with identical entries in site space.
In the transversal channel one simply has

\begin{equation}
    \chi_{xx} (q) = \left( 1 - \frac{1}{2} \Gamma_{xx} \chi_{xx}^{(1)} (q) \right)^{-1} \chi_{xx}^{(1)} (q) 
\end{equation}
and the same for the y-spin direction. The spin rotation symmetry around the z-axis causes the coupling between $S^x$ and $S^y$ to vanish and hence $\chi_{xy} = \chi_{xy} = 0$, $\chi_{xx} = \chi_{yy}$ and $\Gamma_{xx} = \Gamma_{yy}$. By analogy with Ref.
\onlinecite{vilk1997}, a Hartree-Fock decoupling and consistency with $\frac{1}{2} \mathrm{Tr}\Sigma G= \sum_a U\braket{n_{a \us } n_{a \ds}}$ leads to the following ansatz equation~\cite{lessnichtbp} 

\begin{equation}
    \Gamma^a_{xx} = U \frac{\braket{n_{a \us } n_{a \ds} }}{\braket{n_{a \us}} \braket{n_{a \ds}}},
\end{equation}
that is valid for half-filling and the hole doped case, to relate vertex elements and double occupancies.
Together with the sum-rules 

\begin{align}
    \frac{T}{N} \sum_q \chi_{cc}^{aa} (q) &= \braket{n_{a}} + 2 \braket{n_{a \us} n_{a \ds} } - \braket{n_{a}}^2, \label{chsumrule} \\
    \frac{T}{N} \sum_q \chi_{\alpha \alpha}^{aa} (q) &= \braket{n_{a}} - 2 \braket{n_{a \us} n_{a \ds} } \label{spsumrule},
\end{align}
where $\alpha = x,y,z $. 
It is possible to determine all vertex elements self-consistently, first in the transversal channel then in the longitudinal one. 
The self-energy is calculated via 

\begin{align}
\Sigma^{(2)ab} _{\sigma} (k) &= U \delta_{ab}  \braket{n_{a -\sigma}} + \frac{U}{8} \frac{T}{N} \sum_{q} G^{(1)ab}_{\sigma } (k+q) V^{ab}_{l \sigma }(q) \nonumber \\
&~~+ \frac{U}{8} \frac{T}{N} \sum_{q} G^{(1)ab}_{-\sigma } (k+q) V^{ab}_{t \sigma }(q),  
\label{Sigma1}
\end{align}
where the notation $k\equiv (i\omega_n,\mathbf{k})$ is used to represent both the fermionic Matsubara frequency $\omega_n=(2n+1)\pi T$ associated with the integer $n$ and the wave-vector $\mathbf{k}$.
In Eq. \ref{Sigma1}, we define

\begin{align}
    V^{ab}_{l\sigma }(q) &= \Gamma^{b}_{cc} \chi^{ba}_{cc} (q ) + \Gamma^{b}_{zz} \chi^{ba}_{zz} (q ) - \sigma \frac{1}{2} \big( \Gamma^{b}_{zz} \chi^{ba}_{zc} (q) \nonumber \\
    &~~~+ \Gamma^{b}_{ch} \chi^{ba}_{cz} (q ) +  \chi^{ba}_{zc} (q) \Gamma^{a}_{ch} +  \chi^{ba}_{cz} (q ) \Gamma^{a}_{zz} \big) ,  \\
    V^{ab}_{t \sigma }(q) &= \Gamma^{b}_{xx} \chi^{ba}_{xx} (q ) + \Gamma^{b}_{yy} \chi^{ba}_{yy} (q ).
\end{align}
These expressions are obtained by expanding the four-point correlation function for the self-energy once in the longitudinal and once in the transversal channel, and then taking the average and additionally restoring TR symmetry.

The Matsubara Green's function $G^{(2)}$ is then calculated from the Dyson equation

\begin{equation}
    G^{(2)}_{\sigma} \iwnk = \left( i \omega_n - H_{\sigma } (\kf) + \mu - \Sigma^{(2)}_{\sigma} \iwnk \right)^{-1}.
    \label{eq_g2}
\end{equation}
The TPSC internal consistency check is discussed in Appendix~ \ref{sec:trgsigma}.

\section{Numerical details}
\label{sec:num_details}

We use the sparse-ir library~\cite{Shinaoka2017, Li2020, Wallerberger2023} to represent Green's functions and susceptibilities in an efficient manner. 
The library also allows for efficient Fourier transforms between imaginary times and Matsubara frequencies, as well as an efficient calculation of the sum rules Eq.~\ref{chsumrule} and Eq.~\ref{spsumrule}. Convolutions can be efficiently evaluated using fast Fourier transforms. 
The matrix notation in Eq.~\ref{bsel} allows for an efficient implementation.
The equations~\ref{bsel}, \ref{chsumrule} and \ref{spsumrule} present a multi-dimensional root-finding problem for $\Gamma_{cc}$ and $\Gamma_{zz}$. A good starting guess can be obtained by first neglecting the spin-charge coupling $\chi_{cz}$ and $\chi_{zc}$ and solving the resulting one-dimensional root-finding problems. After that, the full root-finding problem is solved. 
All calculations are performed on a $300 \times 300 $ k-point grid.
A python code for TPSC on the Kane-Mele-Hubbard model, that is capable of calculating the observables presented in this article, can be found in the github repository Ref.~\cite{Lessnich_TPSC_2023}.

\section{Results}
\label{sec:results}

In the following, we present results for the antiferromagnetic spin correlation length, the spin Hall conductivity with and without vertex corrections, and the band gap renormalization. 
We also compute a phase diagram from the spin correlation length.
All results are obtained at half-filling.

\subsection{Spin correlation length}
\label{sec:xi}

\begin{figure}
\centering
\includegraphics[width=\linewidth]{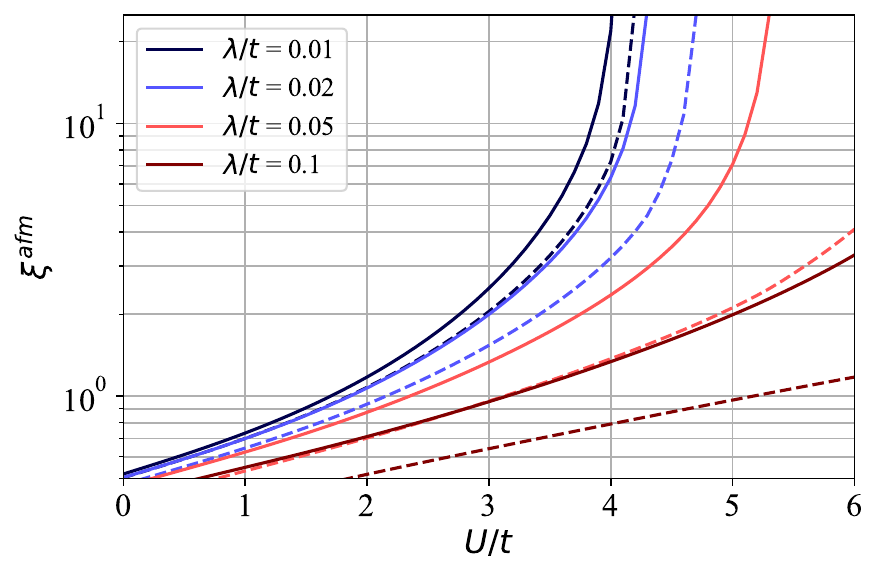}
\caption{Antiferromagnetic spin correlation length $\xi_{\alpha}^{\nk{afm}}$ as a function of $U/t$ for spins aligned in the xy-plane (solid lines) and in the z-direction (dashed lines) at $T/t = 0.001$ and various values of SOC $\lambda / t$. }
\label{fig:xi}
\end{figure}
With an increasing interaction strength $U$, antiferromagnetic spin fluctuations are expected to be larger than the ferromagnetic fluctuations or charge fluctuations. 
Hence, we focus on the antiferromagnetic susceptibilities which are defined as

\begin{equation}
    \chi_{\alpha \alpha}^{\nk{afm}}(q) = \chi_{\alpha \alpha }^{11}(q) - \chi_{\alpha \alpha }^{12}(q) - \chi_{\alpha \alpha }^{21}(q) + \chi_{\alpha \alpha }^{22}(q),
\end{equation}
where $\alpha =x,y,z$ label the component of the spin.
From the maximal value of the antiferromagnetic susceptibilities at wave-vector $\qf_{\nk{max}} = (0,0)$ and Matsubara frequency $q_0=0$, we define the antiferromagnetic correlation length as

\begin{equation}
    \xi_{\alpha}^{\nk{afm}} = \frac{1}{| \qf^{HM}_{\alpha}|},
\end{equation}
where $\qf^{HM}_{\alpha} $ is determined from the condition $\chi^{\nk{afm}}(0, \qf^{HM}_{\alpha} ) = \chi^{\nk{afm}} (0, \qf_{\nk{max}})/2 $. Note that $\xi_{x}^{\nk{afm}} = \xi_{y}^{\nk{afm}}$ due to spin-rotation symmetry around the z-axis.

The results are shown in Fig.~\ref{fig:xi}.
The transversal spin correlation lengths $\xi_{x}^{\nk{afm}} $ and  $\xi_{y}^{\nk{afm}} $ becomes exponentially large at lower U values than the longitudinal one $\xi_{z}^{\nk{afm}} $. 
Hence, the system undergoes a transition to an XY antiferromagnet. 
The corresponding divergence of the spin susceptibility indicates that the transition is of second order in TPSC.
We can also observe that increasing SOC by increasing $\lambda/t$ shifts the transition to higher values of $U$, by decreasing the spin fluctuations.

Further, we find that the spin correlation lengths as a function of temperature saturate at temperatures lower than $T/t = 0.01$ except very close to the phase transition, as already observed in Ref.~\cite{arya2015} (not shown). 

\begin{figure}
\centering
\includegraphics[width=\linewidth]{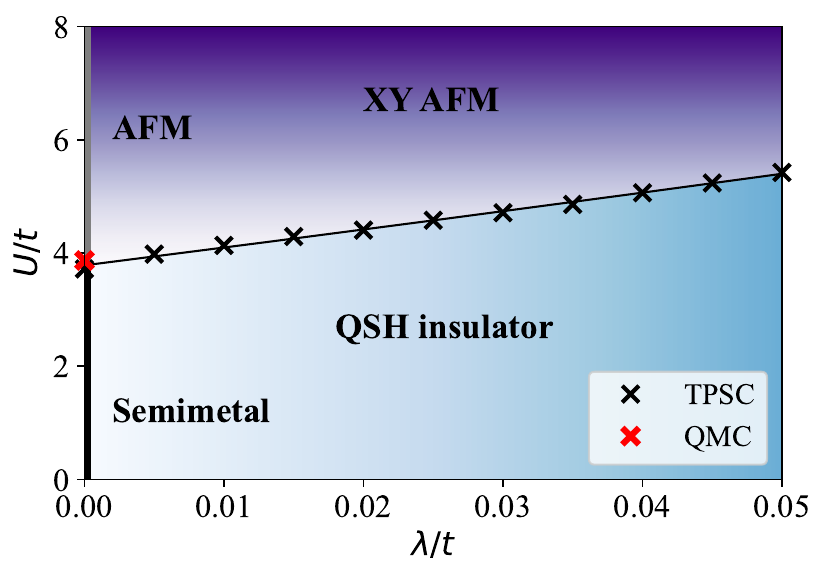}
\caption{Phase diagram for the KMH model. Critical $U_c(\lambda )$ values (black crosses) are obtained from the linear extrapolation of the inverse antiferromagnetic correlation length $1/\xi_x^{afm}$ to $0$ at $T/t = 0.001$. 
The phase transition line is obtained from a quadratic fit. 
We also show the critical $U_c$ (red cross) at zero spin-orbit coupling ($\lambda=0$) for the transition between the semimetal and the antiferromagnet obtained by QMC in Ref.~\onlinecite{Sorella2012}.} 
\label{fig:pd}
\end{figure}

In Fig.~\ref{fig:pd}, we show the $U-\lambda$ phase diagram for the KMH model obtained by estimating the critical value $U_c(\lambda)$ of the phase transition for various $\lambda$ values as the linear extrapolation of $1/\xi_x^{\nk{afm}}$ to $0$ at the lowest temperature we considered ($T/t = 0.001$). 
We only show critical $U$ values for up to $\lambda/t = 0.05$, because for $U/t > 5$ TPSC loses its validity (see Appendix~\ref{sec:trgsigma}). 
The exact location of the phase transition line is not known to the best of our knowledge.
The phase diagram also shows the phase transition between the semimetal and the antiferromagnet at zero SOC, where there is no easy plane for the spins.

Without SOC we obtain $U_c(\lambda = 0 )/t = 3.72$ which agrees well with the previous TPSC result $U_c/t = 3.79 \pm 0.01$~\cite{arya2015} obtained from the zero temperature extrapolation for the crossover to the renormalized classical regime. Our result is also in reasonable good agreement with $U_c/t = 3.869 \pm 0.013$~\cite{Sorella2012} from Quantum Monte Carlo simulations.

\subsection{Spin Hall conductivity}

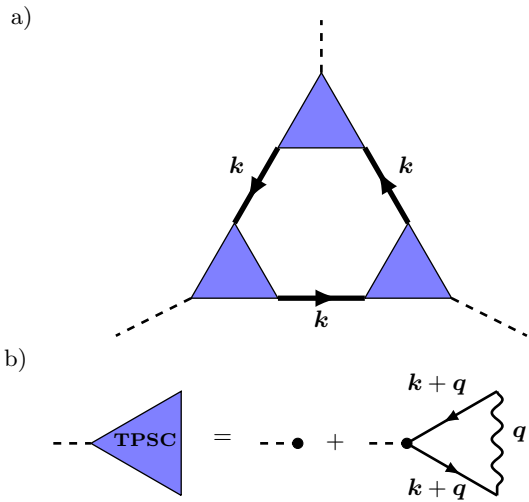
\begin{figure}
\begin{tikzpicture}
  \def\a{2*0.866}
  \node at (-\a * 2.3 , 3. ) [below] {a) };
  \draw [color = black, line width =1, dashed] (\a,-1.) -- (\a+1,-1. -0.5);
  \draw [color = black, line width =1, dashed] (-\a,-1.) -- (-\a-1,-1. -0.5);
  \draw [color = black, line width =1, dashed] (0,2.) -- (0,2.+0.7);
  \node at (-0.65*\a,1. ) [below] {$\kf$};
  \node at (0.65*\a, 1. ) [below] {$\kf$};
  \node at (0 , -1.05 ) [below] {$\kf$};

  \draw [color=black, line width=0.5, fill = blue!50] (-\a, -1.) -- (-\a/3*2,0) -- (-\a/3,-1.) -- (-\a, -1.) ;

  \draw [color=black, line width=0.5, fill = blue!50] (-\a + 4./3. * \a , -1.) -- (-\a/3*2 + 4./3. * \a ,0) -- (-\a/3 + 4./3. * \a ,-1.) -- (-\a + 4./3 * \a, -1.) ;

  \draw [color=black, line width=0.5, fill = blue!50] (-\a + 2./3. * \a , -1.+ 2) -- (-\a/3*2 + 2./3. * \a ,0 + 2) -- (-\a/3 + 2./3. * \a ,-1. + 2 ) -- (-\a + 2./3. * \a , -1. + 2 ) ;

  \begin{scope}[decoration={markings, mark=at position 0.7 with {\arrow{latex}}}] 
  \draw [color=black,line width=2, postaction={decorate}] (-\a/3,-1.) -- (-\a + 4./3. * \a , -1.) ;
  \draw [color=black,line width=2, postaction={decorate}] (-\a/3*2 + 4./3. * \a ,0) 
  -- (-\a/3 + 2./3. * \a ,-1. + 2 );
  \draw [color=black,line width=2, postaction={decorate}] (-\a + 2./3. * \a , -1.+ 2) -- (-\a/3*2,0);
  \end{scope}
\end{tikzpicture}
\begin{tikzpicture}
  \def\a{2*0.866}
  \node at (-1. , 0. ) [below] {b) };
  \def\s{1.2}
  \draw [color = black, line width =1, dashed] (-0.5, -\a/3*2 *\s) -- (0., -\a/3*2 *\s);
  \draw [color=black, line width=0.5, fill = blue!50, rotate = 90] (-\a*\s, -1.*\s) -- (-\a/3*2*\s,0) -- (-\a/3*\s,-1.*\s) -- (-\a*\s, -1.*\s) ;
  \node at (0.6 *\s , -\a/2*\s*1.1 ) [below] {\scriptsize \bf TPSC};
  \node at (1.75 , -\a/2 *\s *1.1 ) [below] {$=$};
  \draw [color = black, line width =1, dashed] (2.25, -\a/3*2 *\s) -- (2.75, -\a/3*2 *\s);
  \draw[black,fill=black] (2.75, -\a/3*2 *\s) circle (.5ex);
  \node at (3.25  , -\a/2 *\s *1.1 ) [below] {$+$};
  \def\x{-4.2}
  \begin{scope}[decoration={markings, mark=at position 0.6 with {\arrow{latex}}}] 
  \draw [color=black, line width=1, rotate = 90, postaction={decorate}] (-\a/3*2*\s  ,0 + \x) -- (-\a*\s , -1.*\s +\x) ;
  \draw [color=black, line width=1, rotate = 90, postaction={decorate}] (-\a/3*\s  ,-1.*\s + \x ) -- (-\a/3*2*\s  ,0 + \x);
  \end{scope}
  \draw [color=black, line width=1, rotate = 90, decorate, decoration={snake, segment length=10, amplitude=2.}]  (-\a/3*\s  ,-1.*\s + \x ) -- (-\a*\s, -1.*\s + \x) ;
  \draw [color = black, line width =1, dashed] (-\x -0.5, -\a/3*2 *\s) -- (-\x, -\a/3*2 *\s);
  \draw[black,fill=black] (-\x, -\a/3*2 *\s) circle (.5ex);
  \node at (-\x + 1.5, -\a/3 *\s*1.5) [below] {$\qf$};
  \node at (-\x + 0.4, -\a/5 ) [below] {$ \kf+\qf$};
  \node at (-\x + 0.4, -\a/3 * \s * 2.5) [below] {$\kf+\qf$};
\end{tikzpicture}
\caption{a) Graphical representation of the expression for the spin Hall conductivity. The thick lines represent the interacting Green's function, which in TPSC is $G^{(2)}$ given in Eq.~\ref{eq_g2}. The triangles with attached dashed lines represent the renormalized vertices $\partial_\mu G^{-1}(k)$ with $\mu$ standing for either $i \omega_n ,~k_x$ or $ k_y $. b) TPSC expression for the renormalized vertex. The thin lines here represent the (here noninteracting) Green's function $G^{(1)}$. A dot with attached dashed line represents a bare vertex $\partial_\mu G^{(1)-1}(k)$. The wiggled line represents either a spin, a charge or a mixed spin-charge excitation.}
\label{fig:vc}
\end{figure}
Similar to Ref.~\onlinecite{Yoshida2012}, we calculate the spin Hall conductivity by evaluating the following expression valid in the presence of interactions and at finite temperature 

\begin{align}
    \sigma^{\nk{SH}} &= \frac{e^2}{h} \frac{\epsilon_{\mu \nu \rho} T }{12 \pi} \nk{Im} \Bigg(  \sum_{  \omega_n \sigma } \nk{sign} ( \sigma ) \int d \kf \nk{Tr} \Big( G_{\sigma}(k )  \nonumber \\
    &~~\times  \partial_{\mu} G^{-1}_{\sigma}(k ) G_{\sigma}(k ) \partial_{\nu } G^{-1}_{\sigma}(k )  G_{\sigma}(k ) \partial_{\rho} G^{-1}_{\sigma}(k ) \Big) \Bigg) ,
    \label{shc1}
\end{align}
where $\epsilon_{\mu \nu \rho}$ is the totally antisymmetric tensor. 
The $\mu,~\nu$ and  $\rho$ indices are summed over and stand for $i \omega_n, k_x, k_y$. The trace runs over the orbital indices. All matrices in the above equations are expressed in the gauge in momentum space which includes the site positions in the unit cell~\cite{noufrakan2018}.

In the zero temperature limit, the sum over Matsubara frequencies becomes an integral. 
The resulting expression is a topological invariant expressed in terms of the Matsubara Green's function~\cite{Ishikawa1987}, which agrees with the Adler-Bell-Jackiw anomaly of the coefficient of the current correlator. 
Hence, at zero temperature, $\sigma^{\nk{SH}}$ must always be an integer multiple of $2\frac{e^2}{h}$. 

Note that the vertex corrections to the electrical conductivity in TPSC correspond to the analogues of the Maki-Thompson and Aslamasov–Larkin contributions~\cite{bergeron2011}. However, for the (spin) Hall conductivity, the Aslamasov–Larkin contributions cancel because of the antisymmetrization in $x \leftrightarrow y$.
A graphical representation of the expression to calculate $\sigma^{\nk{SH}}$ is shown in Fig.~\ref{fig:vc}. Vertex correction to
the spin Hall conductivity in TPSC arise from the excitation and reabsorption of spin, charge or mixed spin-charge excitations.

To evaluate Eq.~\ref{shc1} within TPSC, we use the TPSC Green's function $G^{(2)}$ (Eq.~\ref{eq_g2}). 
Expressions for momentum and frequency derivatives of the inverse Green's function $G^{(2)}(k)^{-1} = i\omega_n - H(\kf) + \mu - \Sigma^{(2)} (k) $ can be found analytically by noticing that when deriving $\Sigma^{(2)} (k)$ given by Eq.~\ref{Sigma1}, the derivatives only act on $G^{(1)}$. Using the identity $\partial_\mu G^{(1)}(k) = - G^{(1)}(k)\partial_\mu G^{(1)-1}(k) G^{(1)}(k) $ the resulting expression can be evaluated numerically. We note that the momentum-dependent vertex corrections show sharp point-like features at $\f{K}$ and $\f{K}'$ near (and beyond) the critical $U_c$ which require a dense k-point grid to resolve.

\begin{figure}
\centering
\includegraphics[width=\linewidth]{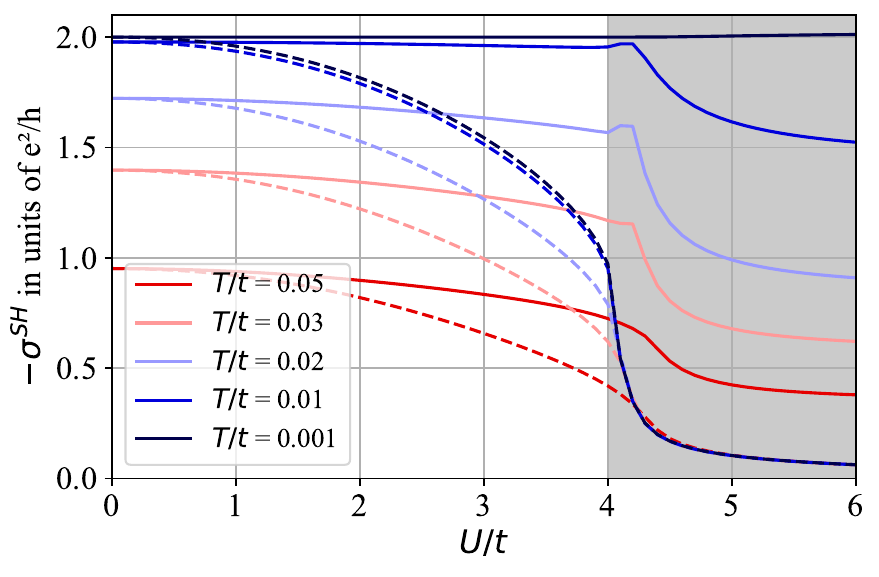}
\caption{Spin Hall conductivity as a function of $U/t$ calculated from Eq.~\ref{shc1} at $\lambda/t = 0.01$ and for different temperatures including momentum-dependent vertex corrections $\partial_{k_\alpha} \Sigma(k)$ (solid) and neglecting them (dashed).
In the vicinity of the phase transition to the XY antiferromagnet at $U/t \approx 4$ the inclusion of vertex corrections almost doubles $\sigma^{SH}$ at all temperatures.
In the greyed out region TPSC is strictly not valid anymore, because the spin correlation length becomes exponentially large in this region.} 
\label{fig:SHC}
\end{figure}
In Fig.~\ref{fig:SHC} we show the results for the calculation of $\sigma^{\nk{SH}}$   with (solid lines) and without (dashed lines) momentum-dependent vertex corrections $\partial_{k_\alpha} \Sigma(k)$. The case without momentum-dependent vertex corrections corresponds to the so-called conductivity bubble. In the greyed out region TPSC is strictly not valid anymore, because the spin correlation length becomes exponentially large in that region.

We first discuss the case without vertex corrections (dashed lines). At all temperatures considered,
the spin Hall conductivity decreases smoothly as the Hubbard interaction increases towards the phase transition. 
At the phase transition, the drop in $\sigma^{\nk{SH}}$ becomes sharper. 
The drop is more pronounced at low temperatures.
In the conductivity bubble, the main effect of the interactions is to broaden the interacting Green's function.
This causes the phase transition to be visible in $\sigma^{\nk{SH}}$ with the conductivity bubble through a drop off.
Without vertex corrections, the quantized value of $\sigma^{\nk{SH}}$ is only reached in the limit where the Hubbard interaction goes to $0$.

We now turn our attention to the case with vertex corrections (solid lines).
The spin Hall conductivity once again decreases with increasing $U$.
However, this decrease is much slower than for the case without vertex corrections.
With vertex corrections, the SHC converges to the quantized value of $-2 e^2/h$ in the zero temperature limit at all values of $U$ below the phase transition.
This is in stark contrast to the case without vertex corrections.
At larger temperatures, where there is no quantization, momentum-dependent vertex corrections also give a large contribution to $\sigma^{\nk{SH}}$.
In the vicinity of the phase transition to the XY antiferromagnet at $U/t \approx 4$, the inclusion of vertex corrections almost doubles the value of $\sigma^{SH}$ at all temperatures.

The impact of the vertex corrections can be explained by the following: When approaching the phase transition, the antiferromagnetic spin fluctuations become strong and scattering of electrons on them yields a large contribution to $\sigma^{\nk{SH}}$.
Numerically, we see that the momentum dependence of the self-energy is rather strong near the phase transition.
This shows that especially in the vicinity of the phase transition, where antiferromagnetic spin fluctuations are strong, using only the conductivity bubble is insufficient and vertex corrections become important.

We also note that, at the phase transition, the decrease in $\sigma^{\nk{SH}}$ is visible through a much smaller kink.
The reason is that, through the enforcement of TR symmetry in the above TPSC self-consistency equations, the system cannot become magnetic even though the spin correlation length becomes exponentially large.
The small peak seen at certain temperatures in the greyed out region, where TPSC is not valid, could be caused by the averaging of the self-energy expressions expanded in the longitudinal and the transversal channels.
Indeed, as discussed in section~\ref{sec:xi}, with SOC the antiferromagnetic instability is reached first for transversal spin fluctuations and only at larger $U$ values for longitudinal ones.

\begin{figure}
\centering
\includegraphics[width=\linewidth]{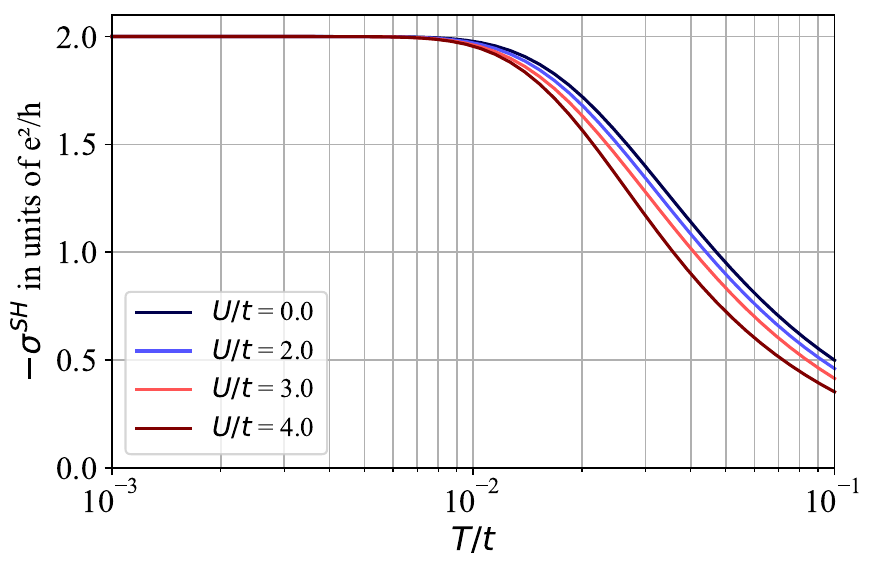}
\caption{Spin Hall conductivity including vertex corrections as a function of $T/t$ calculated from Eq.~\ref{shc1} at $\lambda/t = 0.01$ and for different values of the Hubbard $U$. } 
\label{fig:SHCvsT}
\end{figure}
In Fig.~\ref{fig:SHCvsT}, we show the temperature dependence of $\sigma^{\nk{SH}}$ including vertex corrections for different values of $U$. 
At low temperatures, $\sigma^{\nk{SH}}$ goes to the quantized values of $-2 e^2/h$. 
Deviations from the quantized value start when the temperature becomes of the order of the band gap, which is determined by the value of the SOC.
Increasing the Hubbard $U$ decreases $\sigma^{\nk{SH}}$ but the overall trend stays the same. We conclude that interactions destabilize the QSH state. 
The decrease is explained with a renormalization of the band gap by interactions, which increases the effective temperature in the system.
Hence, the spin Hall conductivity is decreased because more electrons occupy states above the band gap and give canceling contributions to it. 

In the case of the anomalous Hall conductivity, Ref.~\cite{Markov2019} attributes the decrease to a correlation-induced increase of spectral weight within the gap at finite temperatures.
For the (spin) Hall conductivity, 
Refs.~\onlinecite{Yoshida2012,Markov2019} focus mainly on the topological insulator to Mott transition, which in the KMH model is overshadowed by the transition to the XY antiferromagnet. The suppression of $\sigma^{\nk{SH}}$ by interactions at finite temperatures has also been observed in DMFT calculations of the Bernevig-Hughes-Zhang-Hubbard model~\cite{Yoshida2012} and similarly for the anomalous Hall conductivity in the Hubbard model in the presence of a magnetic field~\cite{Markov2019}.

\subsection{Band gap renormalization}

\begin{figure}
\centering
\includegraphics[width=\linewidth]{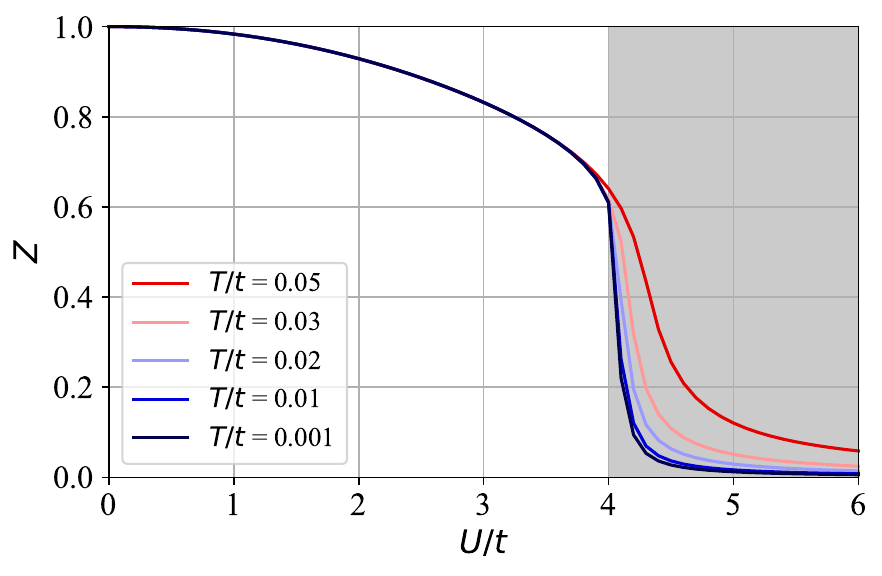}
\caption{Band renormalization $Z$ as a function of $U/t$ for different temperatures and $\lambda / t = 0.01$. In the greyed out region TPSC is not valid anymore. At the phase transition we observe a sharp kink. The temperature dependence is very weak. This indicates that the stronger temperature dependence of $\sigma^{\nk{SH}}$ for higher values of $U$ stems mostly from an effective increase of temperature by renormalizing the gap.}
\label{fig:Z}
\end{figure}
In the following, we discuss how interactions renormalize the band gap and thus affect $\sigma^{\nk{SH}}$.
The quasiparticle weight is defined as 

\begin{equation}
    Z(\omega, \kf ) = \left( 1- \partial_{\omega} \Sigma^{\nk{R}}(\omega , \kf  ) \right)^{-1},
    \label{defZ}
\end{equation}
where $\Sigma^{\nk{R}}(\omega , \kf  )$ is the retarded self-energy on the real frequency axis, indicated by $\omega$.
At $\kf = \Kf,~\Kf' $ the TPSC self-energy $\Sigma^{(2)} (i \omega_n , \kf )$ is diagonal with identical entries that are purely imaginary, therefore $Z(0, \Kf )$ becomes just a number $Z$, which in the following we call the band gap renormalization. 
That interactions renormalize the band gap $\Delta E_g$ by a factor of $Z$ can be seen by expanding the self-energy linearly

\begin{align}
    G_{\pm }( \omega , \Kf ) &=\left( \omega + i \eta \pm \frac{\Delta E_g}{2}  + \mu - \Sigma ( \omega, \Kf )  \right)^{-1} \nonumber \\
    &\approx \left( (1- \partial_\omega \Sigma ( \omega , \Kf )|_{\omega = 0} ) \omega \pm \frac{\Delta E_g}{2} + \mu \right)^{-1} \nonumber \\
    &=  \left( \omega \pm Z \frac{\Delta E_g}{2} + Z \mu \right)^{-1} Z .
\end{align}
For small SOC, the main contributions to $\sigma^{\nk{SH}}$ are localized around $\Kf$ and $\Kf'$, as can be seen in the noninteracting case. At nonzero temperature, the band gap renormalization is crucial for $\sigma^{\nk{SH}}$ since the occupied states above the gap yield canceling contributions to it. Renormalizing the gap hence can be seen as increasing the effective temperature in the system. 

In Fig.~\ref{fig:Z} we show the band gap renormalization $Z$ obtained from our TPSC calculations as a function of $U$. To calculate $Z$ numerically, the analytical continuation to real frequencies is performed using $\partial_\omega \nk{Re} \Sigma^{\nk{R}} (  \omega , \Kf )|_{\omega = 0} \approx  \nk{Im} \Sigma^{(2)} ( i \omega_0 , \Kf ) / \omega_0$, where $\omega_0=\pi T$ is the lowest fermionic Matsubara frequency. 
We observe a sharp drop at the phase transition. 
However, in contrast to $\sigma^{\nk{SH}}$, $Z$ is almost temperature independent below the phase transition. 
This indicates that the stronger temperature dependence of the spin Hall conductivity at larger values of $U$ stems mostly from an effective increase of temperature by renormalizing the gap.
We note that the band gap renormalization and the corresponding decrease of the spin Hall conductivity at finite temperature was also observed in Ref.~\cite{Yoshida2012}. 

In the following we show that the antiferromagnetic spin fluctuations renormalize the band gap.
We consider the diagonal elements of the self-energy at zero frequency and $\kf = \f{K}$ that are responsible for the band gap renormalization.
Since the antiferromagnetic spin fluctuations dominate over charge and ferromagnetic spin fluctuations in the system, especially at large $U$ values, we only keep them. 
Furthermore, from $\chi^{11}_{\alpha \alpha}(q) = \chi^{22}_{\alpha \alpha}(q) $ we have $\chi^{aa}_{\alpha \alpha}(q) \sim \chi_{\alpha \alpha}^{\nk{afm}}(q)/2 $. The Hartee term can be absorbed in the chemical potential.
Focusing only on $\chi_{xx}$ (the same holds for $\chi_{yy}$ and $\chi_{zz}$) we can write 

\begin{align}
\Sigma^{(2)aa} _{\sigma} (i\omega_0,\f{K}) &\sim \frac{U}{16} \frac{T}{N} \sum_{iq_m, \qf \alpha}  G^{(1)aa}_{-\sigma } (i\omega_0 + i\qf_m,\f{K} + \qf )  \nonumber \\
&~~~\times \Gamma^{a}_{xx} \chi^{\nk{afm}}_{xx}(i q_m, \qf ), 
\end{align}
where we see that the leading contribution to the self-energy comes from antiferromagnetic spin fluctuations.

On first sight it might be counter-intuitive that interactions make the band gap smaller, since the common picture in mind is that increasing interactions lead to a formation of Hubbard bands and thus increasing the band gap. 
However, a renormalization of the band structure accompanied by an increase of the effective electron mass through interactions is a common feature in correlated systems.

\section{Conclusion}
\label{sec:conclusion}

In this work, we studied the properties of the
Kane-Mele-Hubbard model at finite temperature with TPSC, which we extended to include SOC.

We calculated the spin Hall conductivity for different values of the Hubbard interaction strength $U$ and of the temperature by using, on the one hand, the conductivity bubble and, on the other hand, by including vertex corrections.
Vertex corrections for the spin Hall conductivity here correspond to the analogues of the Maki-Thompson contributions, describing the excitation and reabsorption of a spin, charge or mixed spin-charge excitation by an electron. 
Our results show that vertex corrections play a crucial role since they are necessary to obtain the quantized value of $-2 e^2 / h$ at $T=0$ and to keep the spin Hall effect sizable at finite temperature.
In fact, in the vicinity of the phase transition to the XY antiferromagnet, the inclusion of vertex corrections almost doubles $\sigma^{SH}$ at all temperatures.

We have also shown that at finite temperature the spin Hall conductivity is reduced by increasing $U$ in a way that can be mostly explained by a renormalization of the gap by spin fluctuations. 
The gap renormalization is stronger when interactions are stronger.

As a function of SOC and $U$, we have also calculated the TPSC phase diagram where we have determined the separation between a XY antiferromagnetic phase at large U values from a spin-Hall insulating phase at small to intermediate U values.

Our main contribution is to show that nonlocal correlation effects that result in a strong momentum dependence of the self-energy can play an important role in the calculation of the spin Hall conductivity at finite temperature and that vertex corrections are extremely important to obtain the zero-temperature quantized value that corresponds to a topological invariant of the Matsubara Green's function. 
TPSC is well suited to study these effects quantitatively in the weak to intermediate coupling regime.

\acknowledgments{
We would like to thank Nicolas Martin, Camille Lahaie, Moïse Rousseau, Abhishek Kumar, Jonas Hauck, Valentin Crépel, Axel Fünfhaus, Falko Pientka, Karim Zantout, Steffen Backes, Giorgio Sangiovanni and Peizhi Mai for helpful discussions.
We acknowledge support by the Deutsche Forschungsgemeinschaft (DFG, German Research Foundation) for funding through TRR 288 – 422213477 (project B05), by the Natural Sciences and Engineering Council of Canada (NSERC) for funding through RGPIN-2019-05312, by the Canada First Research Excellence Fund and by a Vanier Scholarship from NSERC. 
DL is grateful for the hospitality of Institut quantique in Sherbrooke where this work was undertaken.
}

\appendix

\section{$\nk{Tr} ( \Sigma G )$ consistency check}
\label{sec:trgsigma}

\begin{figure}
\centering
\includegraphics[width=\linewidth]{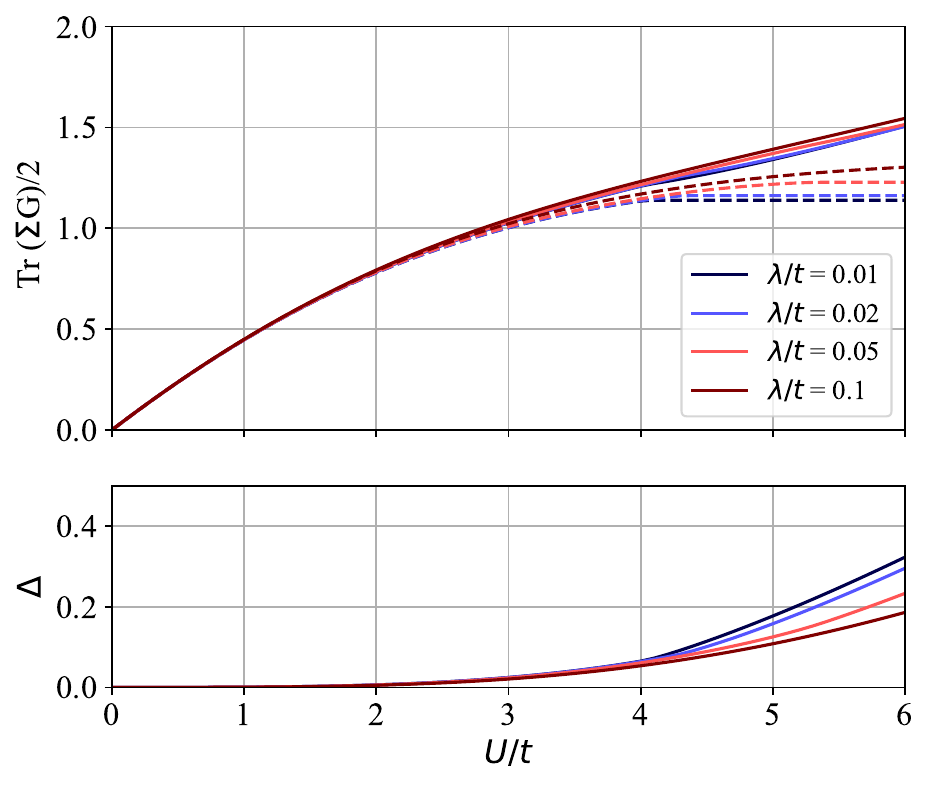}
\caption{$\nk{Tr} \left(  \Sigma G \right)$ consistency check as a function of $U/t$ for different values of SOC and $T/ t = 0.001$. Solid lines are the interacting Green's function $G = G^{(2)}$ and dashed lines the noninteracting one $G = G^{(1)}$. Upper panel: Absolute values of $ \frac{1}{2} \nk{Tr} \left(  \Sigma G \right)$. Lower panel: Relative error $\Delta$ (see eq.~\ref{defdelta}). The deviation is modest up to the phase transition or $U/t \approx 5$ where TPSC starts losing its validity. }
\label{fig:trSigG}
\end{figure}
For the TPSC self-energy $\Sigma^{(2)}$ the following relation between one- and two-particle quantities holds exactly

\begin{align}
    \frac{1}{2} \nk{Tr} \left(  \Sigma^{(2)} G^{(1)} \right) &= \frac{1}{2} \frac{T}{N} \sum_{k a b \sigma}  \Sigma_{\sigma}^{(2)ab}(k) G_{\sigma}^{(1)ba}(k) e^{-i \omega_n 0^-} \nonumber \\
    &= \sum_a U \braket{n_{a\us} n_{a \ds}},
\end{align}
where here the trace runs over orbital and spin indices.
Using $G^{(2)}$ instead of $G^{(1)}$ this becomes the Migdal-Galitskii equation that is an exact relation that does not hold in TPSC but that can be used as an internal consistency check.

In Fig.~\ref{fig:trSigG} we show the following test for different values of SOC. In the upper panel we present the absolute values and in the lower panel the relative derivation

\begin{equation}
    \Delta = \frac{\nk{Tr} \left(  \Sigma^{(2)} G^{(2)} \right) - \nk{Tr} \left(  \Sigma^{(2)} G^{(1)} \right)}{\nk{Tr} \left( \Sigma^{(2)} G^{(1)} \right)}.
    \label{defdelta}
\end{equation}

The error in general gets worse for larger values of $U$.
For low SOC, the deviation is relatively small up to the phase transition, then it starts to diverge rapidly. 
For larger values of SOC, where the phase transition is at larger $U$ values, the error can already be large at the point of the phase transition. 
This indicates that TPSC is not valid anymore after the phase transition or at large enough U values in general. 
The reason is that TPSC is a weak to intermediate coupling method that cannot describe strong coupling physics.
The plateau in the dashed curves involving $G^{(1)}$ corresponds to the vertex $\Gamma_{xx}$ converging to a value, where the spin susceptibility otherwise would have a pole if $\Gamma_{xx}$ would become larger. 
This is another indication that TPSC is not valid anymore beyond the phase transition.

\bibliography{main}

\begin{thebibliography}{82}%
\makeatletter
\providecommand \@ifxundefined [1]{%
 \@ifx{#1\undefined}
}%
\providecommand \@ifnum [1]{%
 \ifnum #1\expandafter \@firstoftwo
 \else \expandafter \@secondoftwo
 \fi
}%
\providecommand \@ifx [1]{%
 \ifx #1\expandafter \@firstoftwo
 \else \expandafter \@secondoftwo
 \fi
}%
\providecommand \natexlab [1]{#1}%
\providecommand \enquote  [1]{``#1''}%
\providecommand \bibnamefont  [1]{#1}%
\providecommand \bibfnamefont [1]{#1}%
\providecommand \citenamefont [1]{#1}%
\providecommand \href@noop [0]{\@secondoftwo}%
\providecommand \href [0]{\begingroup \@sanitize@url \@href}%
\providecommand \@href[1]{\@@startlink{#1}\@@href}%
\providecommand \@@href[1]{\endgroup#1\@@endlink}%
\providecommand \@sanitize@url [0]{\catcode `\\12\catcode `\$12\catcode
  `\&12\catcode `\#12\catcode `\^12\catcode `\_12\catcode `\%12\relax}%
\providecommand \@@startlink[1]{}%
\providecommand \@@endlink[0]{}%
\providecommand \url  [0]{\begingroup\@sanitize@url \@url }%
\providecommand \@url [1]{\endgroup\@href {#1}{\urlprefix }}%
\providecommand \urlprefix  [0]{URL }%
\providecommand \Eprint [0]{\href }%
\providecommand \doibase [0]{https://doi.org/}%
\providecommand \selectlanguage [0]{\@gobble}%
\providecommand \bibinfo  [0]{\@secondoftwo}%
\providecommand \bibfield  [0]{\@secondoftwo}%
\providecommand \translation [1]{[#1]}%
\providecommand \BibitemOpen [0]{}%
\providecommand \bibitemStop [0]{}%
\providecommand \bibitemNoStop [0]{.\EOS\space}%
\providecommand \EOS [0]{\spacefactor3000\relax}%
\providecommand \BibitemShut  [1]{\csname bibitem#1\endcsname}%
\let\auto@bib@innerbib\@empty
\bibitem [{\citenamefont {{D'Yakonov}}\ and\ \citenamefont
  {{Perel'}}(1971)}]{Yakonov71}%
  \BibitemOpen
  \bibfield  {author} {\bibinfo {author} {\bibfnamefont {M.~I.}\ \bibnamefont
  {{D'Yakonov}}}\ and\ \bibinfo {author} {\bibfnamefont {V.~I.}\ \bibnamefont
  {{Perel'}}},\ }\bibfield  {title} {\bibinfo {title} {{Possibility of
  Orienting Electron Spins with Current}},\ }\href@noop {} {\bibfield
  {journal} {\bibinfo  {journal} {Soviet Journal of Experimental and
  Theoretical Physics Letters}\ }\textbf {\bibinfo {volume} {13}},\ \bibinfo
  {pages} {467} (\bibinfo {year} {1971})}\BibitemShut {NoStop}%
\bibitem [{\citenamefont {Kato}\ \emph {et~al.}(2004)\citenamefont {Kato},
  \citenamefont {Myers}, \citenamefont {Gossard},\ and\ \citenamefont
  {Awschalom}}]{Kato2004}%
  \BibitemOpen
  \bibfield  {author} {\bibinfo {author} {\bibfnamefont {Y.~K.}\ \bibnamefont
  {Kato}}, \bibinfo {author} {\bibfnamefont {R.~C.}\ \bibnamefont {Myers}},
  \bibinfo {author} {\bibfnamefont {A.~C.}\ \bibnamefont {Gossard}},\ and\
  \bibinfo {author} {\bibfnamefont {D.~D.}\ \bibnamefont {Awschalom}},\
  }\bibfield  {title} {\bibinfo {title} {Observation of the spin hall effect in
  semiconductors},\ }\href {https://doi.org/10.1126/science.1105514} {\bibfield
   {journal} {\bibinfo  {journal} {Science}\ }\textbf {\bibinfo {volume}
  {306}},\ \bibinfo {pages} {1910} (\bibinfo {year} {2004})},\ \Eprint
  {https://arxiv.org/abs/https://www.science.org/doi/pdf/10.1126/science.1105514}
  {https://www.science.org/doi/pdf/10.1126/science.1105514} \BibitemShut
  {NoStop}%
\bibitem [{\citenamefont {Wunderlich}\ \emph {et~al.}(2005)\citenamefont
  {Wunderlich}, \citenamefont {Kaestner}, \citenamefont {Sinova},\ and\
  \citenamefont {Jungwirth}}]{wunderlich2005}%
  \BibitemOpen
  \bibfield  {author} {\bibinfo {author} {\bibfnamefont {J.}~\bibnamefont
  {Wunderlich}}, \bibinfo {author} {\bibfnamefont {B.}~\bibnamefont
  {Kaestner}}, \bibinfo {author} {\bibfnamefont {J.}~\bibnamefont {Sinova}},\
  and\ \bibinfo {author} {\bibfnamefont {T.}~\bibnamefont {Jungwirth}},\
  }\bibfield  {title} {\bibinfo {title} {Experimental observation of the
  spin-hall effect in a two-dimensional spin-orbit coupled semiconductor
  system},\ }\href {https://doi.org/10.1103/PhysRevLett.94.047204} {\bibfield
  {journal} {\bibinfo  {journal} {Phys. Rev. Lett.}\ }\textbf {\bibinfo
  {volume} {94}},\ \bibinfo {pages} {047204} (\bibinfo {year}
  {2005})}\BibitemShut {NoStop}%
\bibitem [{\citenamefont {Hosten}\ and\ \citenamefont
  {Kwiat}(2008)}]{hosten2008}%
  \BibitemOpen
  \bibfield  {author} {\bibinfo {author} {\bibfnamefont {O.}~\bibnamefont
  {Hosten}}\ and\ \bibinfo {author} {\bibfnamefont {P.}~\bibnamefont {Kwiat}},\
  }\bibfield  {title} {\bibinfo {title} {Observation of the spin hall effect of
  light via weak measurements},\ }\href
  {https://doi.org/10.1126/science.1152697} {\bibfield  {journal} {\bibinfo
  {journal} {Science}\ }\textbf {\bibinfo {volume} {319}},\ \bibinfo {pages}
  {787} (\bibinfo {year} {2008})},\ \Eprint
  {https://arxiv.org/abs/https://www.science.org/doi/pdf/10.1126/science.1152697}
  {https://www.science.org/doi/pdf/10.1126/science.1152697} \BibitemShut
  {NoStop}%
\bibitem [{\citenamefont {Beeler}\ \emph {et~al.}(2013)\citenamefont {Beeler},
  \citenamefont {Williams}, \citenamefont {Jim{\'e}nez-Garc{\'i}a},
  \citenamefont {LeBlanc}, \citenamefont {Perry},\ and\ \citenamefont
  {Spielman}}]{Beeler2013}%
  \BibitemOpen
  \bibfield  {author} {\bibinfo {author} {\bibfnamefont {M.~C.}\ \bibnamefont
  {Beeler}}, \bibinfo {author} {\bibfnamefont {R.~A.}\ \bibnamefont
  {Williams}}, \bibinfo {author} {\bibfnamefont {K.}~\bibnamefont
  {Jim{\'e}nez-Garc{\'i}a}}, \bibinfo {author} {\bibfnamefont {L.~J.}\
  \bibnamefont {LeBlanc}}, \bibinfo {author} {\bibfnamefont {A.~R.}\
  \bibnamefont {Perry}},\ and\ \bibinfo {author} {\bibfnamefont {I.~B.}\
  \bibnamefont {Spielman}},\ }\bibfield  {title} {\bibinfo {title} {The spin
  hall effect in a quantum gas},\ }\href {https://doi.org/10.1038/nature12185}
  {\bibfield  {journal} {\bibinfo  {journal} {Nature}\ }\textbf {\bibinfo
  {volume} {498}},\ \bibinfo {pages} {201} (\bibinfo {year}
  {2013})}\BibitemShut {NoStop}%
\bibitem [{\citenamefont {Aidelsburger}\ \emph {et~al.}(2013)\citenamefont
  {Aidelsburger}, \citenamefont {Atala}, \citenamefont {Lohse}, \citenamefont
  {Barreiro}, \citenamefont {Paredes},\ and\ \citenamefont
  {Bloch}}]{Aidelsburger2013}%
  \BibitemOpen
  \bibfield  {author} {\bibinfo {author} {\bibfnamefont {M.}~\bibnamefont
  {Aidelsburger}}, \bibinfo {author} {\bibfnamefont {M.}~\bibnamefont {Atala}},
  \bibinfo {author} {\bibfnamefont {M.}~\bibnamefont {Lohse}}, \bibinfo
  {author} {\bibfnamefont {J.~T.}\ \bibnamefont {Barreiro}}, \bibinfo {author}
  {\bibfnamefont {B.}~\bibnamefont {Paredes}},\ and\ \bibinfo {author}
  {\bibfnamefont {I.}~\bibnamefont {Bloch}},\ }\bibfield  {title} {\bibinfo
  {title} {Realization of the hofstadter hamiltonian with ultracold atoms in
  optical lattices},\ }\href {https://doi.org/10.1103/PhysRevLett.111.185301}
  {\bibfield  {journal} {\bibinfo  {journal} {Phys. Rev. Lett.}\ }\textbf
  {\bibinfo {volume} {111}},\ \bibinfo {pages} {185301} (\bibinfo {year}
  {2013})}\BibitemShut {NoStop}%
\bibitem [{\citenamefont {Kane}\ and\ \citenamefont
  {Mele}(2005{\natexlab{a}})}]{Kane2005a}%
  \BibitemOpen
  \bibfield  {author} {\bibinfo {author} {\bibfnamefont {C.~L.}\ \bibnamefont
  {Kane}}\ and\ \bibinfo {author} {\bibfnamefont {E.~J.}\ \bibnamefont
  {Mele}},\ }\bibfield  {title} {\bibinfo {title} {${Z}_{2}$ topological order
  and the quantum spin hall effect},\ }\href
  {https://doi.org/10.1103/PhysRevLett.95.146802} {\bibfield  {journal}
  {\bibinfo  {journal} {Phys. Rev. Lett.}\ }\textbf {\bibinfo {volume} {95}},\
  \bibinfo {pages} {146802} (\bibinfo {year} {2005}{\natexlab{a}})}\BibitemShut
  {NoStop}%
\bibitem [{\citenamefont {Kane}\ and\ \citenamefont
  {Mele}(2005{\natexlab{b}})}]{Kane2005b}%
  \BibitemOpen
  \bibfield  {author} {\bibinfo {author} {\bibfnamefont {C.~L.}\ \bibnamefont
  {Kane}}\ and\ \bibinfo {author} {\bibfnamefont {E.~J.}\ \bibnamefont
  {Mele}},\ }\bibfield  {title} {\bibinfo {title} {Quantum spin hall effect in
  graphene},\ }\href {https://doi.org/10.1103/PhysRevLett.95.226801} {\bibfield
   {journal} {\bibinfo  {journal} {Phys. Rev. Lett.}\ }\textbf {\bibinfo
  {volume} {95}},\ \bibinfo {pages} {226801} (\bibinfo {year}
  {2005}{\natexlab{b}})}\BibitemShut {NoStop}%
\bibitem [{\citenamefont {Sichau}\ \emph {et~al.}(2019)\citenamefont {Sichau},
  \citenamefont {Prada}, \citenamefont {Anlauf}, \citenamefont {Lyon},
  \citenamefont {Bosnjak}, \citenamefont {Tiemann},\ and\ \citenamefont
  {Blick}}]{sichau2019}%
  \BibitemOpen
  \bibfield  {author} {\bibinfo {author} {\bibfnamefont {J.}~\bibnamefont
  {Sichau}}, \bibinfo {author} {\bibfnamefont {M.}~\bibnamefont {Prada}},
  \bibinfo {author} {\bibfnamefont {T.}~\bibnamefont {Anlauf}}, \bibinfo
  {author} {\bibfnamefont {T.~J.}\ \bibnamefont {Lyon}}, \bibinfo {author}
  {\bibfnamefont {B.}~\bibnamefont {Bosnjak}}, \bibinfo {author} {\bibfnamefont
  {L.}~\bibnamefont {Tiemann}},\ and\ \bibinfo {author} {\bibfnamefont {R.~H.}\
  \bibnamefont {Blick}},\ }\bibfield  {title} {\bibinfo {title} {Resonance
  microwave measurements of an intrinsic spin-orbit coupling gap in graphene: A
  possible indication of a topological state},\ }\href
  {https://doi.org/10.1103/PhysRevLett.122.046403} {\bibfield  {journal}
  {\bibinfo  {journal} {Phys. Rev. Lett.}\ }\textbf {\bibinfo {volume} {122}},\
  \bibinfo {pages} {046403} (\bibinfo {year} {2019})}\BibitemShut {NoStop}%
\bibitem [{\citenamefont {Yao}\ \emph {et~al.}(2007)\citenamefont {Yao},
  \citenamefont {Ye}, \citenamefont {Qi}, \citenamefont {Zhang},\ and\
  \citenamefont {Fang}}]{Yao2007}%
  \BibitemOpen
  \bibfield  {author} {\bibinfo {author} {\bibfnamefont {Y.}~\bibnamefont
  {Yao}}, \bibinfo {author} {\bibfnamefont {F.}~\bibnamefont {Ye}}, \bibinfo
  {author} {\bibfnamefont {X.-L.}\ \bibnamefont {Qi}}, \bibinfo {author}
  {\bibfnamefont {S.-C.}\ \bibnamefont {Zhang}},\ and\ \bibinfo {author}
  {\bibfnamefont {Z.}~\bibnamefont {Fang}},\ }\bibfield  {title} {\bibinfo
  {title} {Spin-orbit gap of graphene: First-principles calculations},\ }\href
  {https://doi.org/10.1103/PhysRevB.75.041401} {\bibfield  {journal} {\bibinfo
  {journal} {Phys. Rev. B}\ }\textbf {\bibinfo {volume} {75}},\ \bibinfo
  {pages} {041401} (\bibinfo {year} {2007})}\BibitemShut {NoStop}%
\bibitem [{\citenamefont {Min}\ \emph {et~al.}(2006)\citenamefont {Min},
  \citenamefont {Hill}, \citenamefont {Sinitsyn}, \citenamefont {Sahu},
  \citenamefont {Kleinman},\ and\ \citenamefont {MacDonald}}]{Min2006}%
  \BibitemOpen
  \bibfield  {author} {\bibinfo {author} {\bibfnamefont {H.}~\bibnamefont
  {Min}}, \bibinfo {author} {\bibfnamefont {J.~E.}\ \bibnamefont {Hill}},
  \bibinfo {author} {\bibfnamefont {N.~A.}\ \bibnamefont {Sinitsyn}}, \bibinfo
  {author} {\bibfnamefont {B.~R.}\ \bibnamefont {Sahu}}, \bibinfo {author}
  {\bibfnamefont {L.}~\bibnamefont {Kleinman}},\ and\ \bibinfo {author}
  {\bibfnamefont {A.~H.}\ \bibnamefont {MacDonald}},\ }\bibfield  {title}
  {\bibinfo {title} {Intrinsic and rashba spin-orbit interactions in graphene
  sheets},\ }\href {https://doi.org/10.1103/PhysRevB.74.165310} {\bibfield
  {journal} {\bibinfo  {journal} {Phys. Rev. B}\ }\textbf {\bibinfo {volume}
  {74}},\ \bibinfo {pages} {165310} (\bibinfo {year} {2006})}\BibitemShut
  {NoStop}%
\bibitem [{\citenamefont {Huertas-Hernando}\ \emph {et~al.}(2006)\citenamefont
  {Huertas-Hernando}, \citenamefont {Guinea},\ and\ \citenamefont
  {Brataas}}]{huertas2006}%
  \BibitemOpen
  \bibfield  {author} {\bibinfo {author} {\bibfnamefont {D.}~\bibnamefont
  {Huertas-Hernando}}, \bibinfo {author} {\bibfnamefont {F.}~\bibnamefont
  {Guinea}},\ and\ \bibinfo {author} {\bibfnamefont {A.}~\bibnamefont
  {Brataas}},\ }\bibfield  {title} {\bibinfo {title} {Spin-orbit coupling in
  curved graphene, fullerenes, nanotubes, and nanotube caps},\ }\href
  {https://doi.org/10.1103/PhysRevB.74.155426} {\bibfield  {journal} {\bibinfo
  {journal} {Phys. Rev. B}\ }\textbf {\bibinfo {volume} {74}},\ \bibinfo
  {pages} {155426} (\bibinfo {year} {2006})}\BibitemShut {NoStop}%
\bibitem [{\citenamefont {König}\ \emph {et~al.}(2007)\citenamefont {König},
  \citenamefont {Wiedmann}, \citenamefont {Brüne}, \citenamefont {Roth},
  \citenamefont {Buhmann}, \citenamefont {Molenkamp}, \citenamefont {Qi},\ and\
  \citenamefont {Zhang}}]{konig2007}%
  \BibitemOpen
  \bibfield  {author} {\bibinfo {author} {\bibfnamefont {M.}~\bibnamefont
  {König}}, \bibinfo {author} {\bibfnamefont {S.}~\bibnamefont {Wiedmann}},
  \bibinfo {author} {\bibfnamefont {C.}~\bibnamefont {Brüne}}, \bibinfo
  {author} {\bibfnamefont {A.}~\bibnamefont {Roth}}, \bibinfo {author}
  {\bibfnamefont {H.}~\bibnamefont {Buhmann}}, \bibinfo {author} {\bibfnamefont
  {L.~W.}\ \bibnamefont {Molenkamp}}, \bibinfo {author} {\bibfnamefont {X.-L.}\
  \bibnamefont {Qi}},\ and\ \bibinfo {author} {\bibfnamefont {S.-C.}\
  \bibnamefont {Zhang}},\ }\bibfield  {title} {\bibinfo {title} {Quantum spin
  hall insulator state in hgte quantum wells},\ }\href
  {https://doi.org/10.1126/science.1148047} {\bibfield  {journal} {\bibinfo
  {journal} {Science}\ }\textbf {\bibinfo {volume} {318}},\ \bibinfo {pages}
  {766} (\bibinfo {year} {2007})},\ \Eprint
  {https://arxiv.org/abs/https://www.science.org/doi/pdf/10.1126/science.1148047}
  {https://www.science.org/doi/pdf/10.1126/science.1148047} \BibitemShut
  {NoStop}%
\bibitem [{\citenamefont {Bernevig}\ \emph {et~al.}(2006)\citenamefont
  {Bernevig}, \citenamefont {Hughes},\ and\ \citenamefont
  {Zhang}}]{bernevig2006}%
  \BibitemOpen
  \bibfield  {author} {\bibinfo {author} {\bibfnamefont {B.~A.}\ \bibnamefont
  {Bernevig}}, \bibinfo {author} {\bibfnamefont {T.~L.}\ \bibnamefont
  {Hughes}},\ and\ \bibinfo {author} {\bibfnamefont {S.-C.}\ \bibnamefont
  {Zhang}},\ }\bibfield  {title} {\bibinfo {title} {Quantum spin hall effect
  and topological phase transition in hgte quantum wells},\ }\href
  {https://doi.org/10.1126/science.1133734} {\bibfield  {journal} {\bibinfo
  {journal} {Science}\ }\textbf {\bibinfo {volume} {314}},\ \bibinfo {pages}
  {1757} (\bibinfo {year} {2006})},\ \Eprint
  {https://arxiv.org/abs/https://www.science.org/doi/pdf/10.1126/science.1133734}
  {https://www.science.org/doi/pdf/10.1126/science.1133734} \BibitemShut
  {NoStop}%
\bibitem [{\citenamefont {Roth}\ \emph {et~al.}(2009)\citenamefont {Roth},
  \citenamefont {Brüne}, \citenamefont {Buhmann}, \citenamefont {Molenkamp},
  \citenamefont {Maciejko}, \citenamefont {Qi},\ and\ \citenamefont
  {Zhang}}]{roth2009}%
  \BibitemOpen
  \bibfield  {author} {\bibinfo {author} {\bibfnamefont {A.}~\bibnamefont
  {Roth}}, \bibinfo {author} {\bibfnamefont {C.}~\bibnamefont {Brüne}},
  \bibinfo {author} {\bibfnamefont {H.}~\bibnamefont {Buhmann}}, \bibinfo
  {author} {\bibfnamefont {L.~W.}\ \bibnamefont {Molenkamp}}, \bibinfo {author}
  {\bibfnamefont {J.}~\bibnamefont {Maciejko}}, \bibinfo {author}
  {\bibfnamefont {X.-L.}\ \bibnamefont {Qi}},\ and\ \bibinfo {author}
  {\bibfnamefont {S.-C.}\ \bibnamefont {Zhang}},\ }\bibfield  {title} {\bibinfo
  {title} {Nonlocal transport in the quantum spin hall state},\ }\href
  {https://doi.org/10.1126/science.1174736} {\bibfield  {journal} {\bibinfo
  {journal} {Science}\ }\textbf {\bibinfo {volume} {325}},\ \bibinfo {pages}
  {294} (\bibinfo {year} {2009})},\ \Eprint
  {https://arxiv.org/abs/https://www.science.org/doi/pdf/10.1126/science.1174736}
  {https://www.science.org/doi/pdf/10.1126/science.1174736} \BibitemShut
  {NoStop}%
\bibitem [{\citenamefont {Li}\ \emph {et~al.}(2015)\citenamefont {Li},
  \citenamefont {Wang}, \citenamefont {Fu}, \citenamefont {Du}, \citenamefont
  {Schreiber}, \citenamefont {Mu}, \citenamefont {Liu}, \citenamefont
  {Sullivan}, \citenamefont {Cs\'athy}, \citenamefont {Lin},\ and\
  \citenamefont {Du}}]{tingxin2015}%
  \BibitemOpen
  \bibfield  {author} {\bibinfo {author} {\bibfnamefont {T.}~\bibnamefont
  {Li}}, \bibinfo {author} {\bibfnamefont {P.}~\bibnamefont {Wang}}, \bibinfo
  {author} {\bibfnamefont {H.}~\bibnamefont {Fu}}, \bibinfo {author}
  {\bibfnamefont {L.}~\bibnamefont {Du}}, \bibinfo {author} {\bibfnamefont
  {K.~A.}\ \bibnamefont {Schreiber}}, \bibinfo {author} {\bibfnamefont
  {X.}~\bibnamefont {Mu}}, \bibinfo {author} {\bibfnamefont {X.}~\bibnamefont
  {Liu}}, \bibinfo {author} {\bibfnamefont {G.}~\bibnamefont {Sullivan}},
  \bibinfo {author} {\bibfnamefont {G.~A.}\ \bibnamefont {Cs\'athy}}, \bibinfo
  {author} {\bibfnamefont {X.}~\bibnamefont {Lin}},\ and\ \bibinfo {author}
  {\bibfnamefont {R.-R.}\ \bibnamefont {Du}},\ }\bibfield  {title} {\bibinfo
  {title} {Observation of a helical luttinger liquid in
  $\mathrm{InAs}/\mathrm{GaSb}$ quantum spin hall edges},\ }\href
  {https://doi.org/10.1103/PhysRevLett.115.136804} {\bibfield  {journal}
  {\bibinfo  {journal} {Phys. Rev. Lett.}\ }\textbf {\bibinfo {volume} {115}},\
  \bibinfo {pages} {136804} (\bibinfo {year} {2015})}\BibitemShut {NoStop}%
\bibitem [{\citenamefont {Knez}\ \emph {et~al.}(2011)\citenamefont {Knez},
  \citenamefont {Du},\ and\ \citenamefont {Sullivan}}]{knez2011}%
  \BibitemOpen
  \bibfield  {author} {\bibinfo {author} {\bibfnamefont {I.}~\bibnamefont
  {Knez}}, \bibinfo {author} {\bibfnamefont {R.-R.}\ \bibnamefont {Du}},\ and\
  \bibinfo {author} {\bibfnamefont {G.}~\bibnamefont {Sullivan}},\ }\bibfield
  {title} {\bibinfo {title} {Evidence for helical edge modes in inverted
  $\mathrm{InAs}/\mathrm{GaSb}$ quantum wells},\ }\href
  {https://doi.org/10.1103/PhysRevLett.107.136603} {\bibfield  {journal}
  {\bibinfo  {journal} {Phys. Rev. Lett.}\ }\textbf {\bibinfo {volume} {107}},\
  \bibinfo {pages} {136603} (\bibinfo {year} {2011})}\BibitemShut {NoStop}%
\bibitem [{\citenamefont {Wu}\ \emph {et~al.}(2018)\citenamefont {Wu},
  \citenamefont {Fatemi}, \citenamefont {Gibson}, \citenamefont {Watanabe},
  \citenamefont {Taniguchi}, \citenamefont {Cava},\ and\ \citenamefont
  {Jarillo-Herrero}}]{wu2018}%
  \BibitemOpen
  \bibfield  {author} {\bibinfo {author} {\bibfnamefont {S.}~\bibnamefont
  {Wu}}, \bibinfo {author} {\bibfnamefont {V.}~\bibnamefont {Fatemi}}, \bibinfo
  {author} {\bibfnamefont {Q.~D.}\ \bibnamefont {Gibson}}, \bibinfo {author}
  {\bibfnamefont {K.}~\bibnamefont {Watanabe}}, \bibinfo {author}
  {\bibfnamefont {T.}~\bibnamefont {Taniguchi}}, \bibinfo {author}
  {\bibfnamefont {R.~J.}\ \bibnamefont {Cava}},\ and\ \bibinfo {author}
  {\bibfnamefont {P.}~\bibnamefont {Jarillo-Herrero}},\ }\bibfield  {title}
  {\bibinfo {title} {Observation of the quantum spin hall effect up to 100
  kelvin in a monolayer crystal},\ }\href
  {https://doi.org/10.1126/science.aan6003} {\bibfield  {journal} {\bibinfo
  {journal} {Science}\ }\textbf {\bibinfo {volume} {359}},\ \bibinfo {pages}
  {76} (\bibinfo {year} {2018})},\ \Eprint
  {https://arxiv.org/abs/https://www.science.org/doi/pdf/10.1126/science.aan6003}
  {https://www.science.org/doi/pdf/10.1126/science.aan6003} \BibitemShut
  {NoStop}%
\bibitem [{\citenamefont {Tang}\ \emph {et~al.}(2017)\citenamefont {Tang},
  \citenamefont {Zhang}, \citenamefont {Wong}, \citenamefont {Pedramrazi},
  \citenamefont {Tsai}, \citenamefont {Jia}, \citenamefont {Moritz},
  \citenamefont {Claassen}, \citenamefont {Ryu}, \citenamefont {Kahn},
  \citenamefont {Jiang}, \citenamefont {Yan}, \citenamefont {Hashimoto},
  \citenamefont {Lu}, \citenamefont {Moore}, \citenamefont {Hwang},
  \citenamefont {Hwang}, \citenamefont {Hussain}, \citenamefont {Chen},
  \citenamefont {Ugeda}, \citenamefont {Liu}, \citenamefont {Xie},
  \citenamefont {Devereaux}, \citenamefont {Crommie}, \citenamefont {Mo},\ and\
  \citenamefont {Shen}}]{Tang2017}%
  \BibitemOpen
  \bibfield  {author} {\bibinfo {author} {\bibfnamefont {S.}~\bibnamefont
  {Tang}}, \bibinfo {author} {\bibfnamefont {C.}~\bibnamefont {Zhang}},
  \bibinfo {author} {\bibfnamefont {D.}~\bibnamefont {Wong}}, \bibinfo {author}
  {\bibfnamefont {Z.}~\bibnamefont {Pedramrazi}}, \bibinfo {author}
  {\bibfnamefont {H.-Z.}\ \bibnamefont {Tsai}}, \bibinfo {author}
  {\bibfnamefont {C.}~\bibnamefont {Jia}}, \bibinfo {author} {\bibfnamefont
  {B.}~\bibnamefont {Moritz}}, \bibinfo {author} {\bibfnamefont
  {M.}~\bibnamefont {Claassen}}, \bibinfo {author} {\bibfnamefont
  {H.}~\bibnamefont {Ryu}}, \bibinfo {author} {\bibfnamefont {S.}~\bibnamefont
  {Kahn}}, \bibinfo {author} {\bibfnamefont {J.}~\bibnamefont {Jiang}},
  \bibinfo {author} {\bibfnamefont {H.}~\bibnamefont {Yan}}, \bibinfo {author}
  {\bibfnamefont {M.}~\bibnamefont {Hashimoto}}, \bibinfo {author}
  {\bibfnamefont {D.}~\bibnamefont {Lu}}, \bibinfo {author} {\bibfnamefont
  {R.~G.}\ \bibnamefont {Moore}}, \bibinfo {author} {\bibfnamefont {C.-C.}\
  \bibnamefont {Hwang}}, \bibinfo {author} {\bibfnamefont {C.}~\bibnamefont
  {Hwang}}, \bibinfo {author} {\bibfnamefont {Z.}~\bibnamefont {Hussain}},
  \bibinfo {author} {\bibfnamefont {Y.}~\bibnamefont {Chen}}, \bibinfo {author}
  {\bibfnamefont {M.~M.}\ \bibnamefont {Ugeda}}, \bibinfo {author}
  {\bibfnamefont {Z.}~\bibnamefont {Liu}}, \bibinfo {author} {\bibfnamefont
  {X.}~\bibnamefont {Xie}}, \bibinfo {author} {\bibfnamefont {T.~P.}\
  \bibnamefont {Devereaux}}, \bibinfo {author} {\bibfnamefont {M.~F.}\
  \bibnamefont {Crommie}}, \bibinfo {author} {\bibfnamefont {S.-K.}\
  \bibnamefont {Mo}},\ and\ \bibinfo {author} {\bibfnamefont {Z.-X.}\
  \bibnamefont {Shen}},\ }\bibfield  {title} {\bibinfo {title} {Quantum spin
  hall state in monolayer 1t'-wte2},\ }\href
  {https://doi.org/10.1038/nphys4174} {\bibfield  {journal} {\bibinfo
  {journal} {Nature Physics}\ }\textbf {\bibinfo {volume} {13}},\ \bibinfo
  {pages} {683} (\bibinfo {year} {2017})}\BibitemShut {NoStop}%
\bibitem [{\citenamefont {Fei}\ \emph {et~al.}(2017)\citenamefont {Fei},
  \citenamefont {Palomaki}, \citenamefont {Wu}, \citenamefont {Zhao},
  \citenamefont {Cai}, \citenamefont {Sun}, \citenamefont {Nguyen},
  \citenamefont {Finney}, \citenamefont {Xu},\ and\ \citenamefont
  {Cobden}}]{Fei2017}%
  \BibitemOpen
  \bibfield  {author} {\bibinfo {author} {\bibfnamefont {Z.}~\bibnamefont
  {Fei}}, \bibinfo {author} {\bibfnamefont {T.}~\bibnamefont {Palomaki}},
  \bibinfo {author} {\bibfnamefont {S.}~\bibnamefont {Wu}}, \bibinfo {author}
  {\bibfnamefont {W.}~\bibnamefont {Zhao}}, \bibinfo {author} {\bibfnamefont
  {X.}~\bibnamefont {Cai}}, \bibinfo {author} {\bibfnamefont {B.}~\bibnamefont
  {Sun}}, \bibinfo {author} {\bibfnamefont {P.}~\bibnamefont {Nguyen}},
  \bibinfo {author} {\bibfnamefont {J.}~\bibnamefont {Finney}}, \bibinfo
  {author} {\bibfnamefont {X.}~\bibnamefont {Xu}},\ and\ \bibinfo {author}
  {\bibfnamefont {D.~H.}\ \bibnamefont {Cobden}},\ }\bibfield  {title}
  {\bibinfo {title} {Edge conduction in monolayer wte2},\ }\href
  {https://doi.org/10.1038/nphys4091} {\bibfield  {journal} {\bibinfo
  {journal} {Nature Physics}\ }\textbf {\bibinfo {volume} {13}},\ \bibinfo
  {pages} {677} (\bibinfo {year} {2017})}\BibitemShut {NoStop}%
\bibitem [{\citenamefont {Collins}\ \emph {et~al.}(2018)\citenamefont
  {Collins}, \citenamefont {Tadich}, \citenamefont {Wu}, \citenamefont {Gomes},
  \citenamefont {Rodrigues}, \citenamefont {Liu}, \citenamefont {Hellerstedt},
  \citenamefont {Ryu}, \citenamefont {Tang}, \citenamefont {Mo}, \citenamefont
  {Adam}, \citenamefont {Yang}, \citenamefont {Fuhrer},\ and\ \citenamefont
  {Edmonds}}]{Collins2018}%
  \BibitemOpen
  \bibfield  {author} {\bibinfo {author} {\bibfnamefont {J.~L.}\ \bibnamefont
  {Collins}}, \bibinfo {author} {\bibfnamefont {A.}~\bibnamefont {Tadich}},
  \bibinfo {author} {\bibfnamefont {W.}~\bibnamefont {Wu}}, \bibinfo {author}
  {\bibfnamefont {L.~C.}\ \bibnamefont {Gomes}}, \bibinfo {author}
  {\bibfnamefont {J.~N.~B.}\ \bibnamefont {Rodrigues}}, \bibinfo {author}
  {\bibfnamefont {C.}~\bibnamefont {Liu}}, \bibinfo {author} {\bibfnamefont
  {J.}~\bibnamefont {Hellerstedt}}, \bibinfo {author} {\bibfnamefont
  {H.}~\bibnamefont {Ryu}}, \bibinfo {author} {\bibfnamefont {S.}~\bibnamefont
  {Tang}}, \bibinfo {author} {\bibfnamefont {S.-K.}\ \bibnamefont {Mo}},
  \bibinfo {author} {\bibfnamefont {S.}~\bibnamefont {Adam}}, \bibinfo {author}
  {\bibfnamefont {S.~A.}\ \bibnamefont {Yang}}, \bibinfo {author}
  {\bibfnamefont {M.~S.}\ \bibnamefont {Fuhrer}},\ and\ \bibinfo {author}
  {\bibfnamefont {M.~T.}\ \bibnamefont {Edmonds}},\ }\bibfield  {title}
  {\bibinfo {title} {Electric-field-tuned topological phase transition in
  ultrathin na3bi},\ }\href {https://doi.org/10.1038/s41586-018-0788-5}
  {\bibfield  {journal} {\bibinfo  {journal} {Nature}\ }\textbf {\bibinfo
  {volume} {564}},\ \bibinfo {pages} {390} (\bibinfo {year}
  {2018})}\BibitemShut {NoStop}%
\bibitem [{\citenamefont {Island}\ \emph {et~al.}(2019)\citenamefont {Island},
  \citenamefont {Cui}, \citenamefont {Lewandowski}, \citenamefont {Khoo},
  \citenamefont {Spanton}, \citenamefont {Zhou}, \citenamefont {Rhodes},
  \citenamefont {Hone}, \citenamefont {Taniguchi}, \citenamefont {Watanabe},
  \citenamefont {Levitov}, \citenamefont {Zaletel},\ and\ \citenamefont
  {Young}}]{Island2019}%
  \BibitemOpen
  \bibfield  {author} {\bibinfo {author} {\bibfnamefont {J.~O.}\ \bibnamefont
  {Island}}, \bibinfo {author} {\bibfnamefont {X.}~\bibnamefont {Cui}},
  \bibinfo {author} {\bibfnamefont {C.}~\bibnamefont {Lewandowski}}, \bibinfo
  {author} {\bibfnamefont {J.~Y.}\ \bibnamefont {Khoo}}, \bibinfo {author}
  {\bibfnamefont {E.~M.}\ \bibnamefont {Spanton}}, \bibinfo {author}
  {\bibfnamefont {H.}~\bibnamefont {Zhou}}, \bibinfo {author} {\bibfnamefont
  {D.}~\bibnamefont {Rhodes}}, \bibinfo {author} {\bibfnamefont {J.~C.}\
  \bibnamefont {Hone}}, \bibinfo {author} {\bibfnamefont {T.}~\bibnamefont
  {Taniguchi}}, \bibinfo {author} {\bibfnamefont {K.}~\bibnamefont {Watanabe}},
  \bibinfo {author} {\bibfnamefont {L.~S.}\ \bibnamefont {Levitov}}, \bibinfo
  {author} {\bibfnamefont {M.~P.}\ \bibnamefont {Zaletel}},\ and\ \bibinfo
  {author} {\bibfnamefont {A.~F.}\ \bibnamefont {Young}},\ }\bibfield  {title}
  {\bibinfo {title} {Spin--orbit-driven band inversion in bilayer graphene by
  the van der waals proximity effect},\ }\href
  {https://doi.org/10.1038/s41586-019-1304-2} {\bibfield  {journal} {\bibinfo
  {journal} {Nature}\ }\textbf {\bibinfo {volume} {571}},\ \bibinfo {pages}
  {85} (\bibinfo {year} {2019})}\BibitemShut {NoStop}%
\bibitem [{\citenamefont {Eck}\ \emph {et~al.}(2022)\citenamefont {Eck},
  \citenamefont {Ortix}, \citenamefont {Consiglio}, \citenamefont {Erhardt},
  \citenamefont {Bauernfeind}, \citenamefont {Moser}, \citenamefont {Claessen},
  \citenamefont {Di~Sante},\ and\ \citenamefont {Sangiovanni}}]{eck_2022}%
  \BibitemOpen
  \bibfield  {author} {\bibinfo {author} {\bibfnamefont {P.}~\bibnamefont
  {Eck}}, \bibinfo {author} {\bibfnamefont {C.}~\bibnamefont {Ortix}}, \bibinfo
  {author} {\bibfnamefont {A.}~\bibnamefont {Consiglio}}, \bibinfo {author}
  {\bibfnamefont {J.}~\bibnamefont {Erhardt}}, \bibinfo {author} {\bibfnamefont
  {M.}~\bibnamefont {Bauernfeind}}, \bibinfo {author} {\bibfnamefont
  {S.}~\bibnamefont {Moser}}, \bibinfo {author} {\bibfnamefont
  {R.}~\bibnamefont {Claessen}}, \bibinfo {author} {\bibfnamefont
  {D.}~\bibnamefont {Di~Sante}},\ and\ \bibinfo {author} {\bibfnamefont
  {G.}~\bibnamefont {Sangiovanni}},\ }\bibfield  {title} {\bibinfo {title}
  {Real-space obstruction in quantum spin hall insulators},\ }\href
  {https://doi.org/10.1103/PhysRevB.106.195143} {\bibfield  {journal} {\bibinfo
   {journal} {Phys. Rev. B}\ }\textbf {\bibinfo {volume} {106}},\ \bibinfo
  {pages} {195143} (\bibinfo {year} {2022})}\BibitemShut {NoStop}%
\bibitem [{\citenamefont {Bauernfeind}\ \emph {et~al.}(2021)\citenamefont
  {Bauernfeind}, \citenamefont {Erhardt}, \citenamefont {Eck}, \citenamefont
  {Thakur}, \citenamefont {Gabel}, \citenamefont {Lee}, \citenamefont
  {Sch{\"a}fer}, \citenamefont {Moser}, \citenamefont {Di~Sante}, \citenamefont
  {Claessen},\ and\ \citenamefont {Sangiovanni}}]{Bauernfeind2021}%
  \BibitemOpen
  \bibfield  {author} {\bibinfo {author} {\bibfnamefont {M.}~\bibnamefont
  {Bauernfeind}}, \bibinfo {author} {\bibfnamefont {J.}~\bibnamefont
  {Erhardt}}, \bibinfo {author} {\bibfnamefont {P.}~\bibnamefont {Eck}},
  \bibinfo {author} {\bibfnamefont {P.~K.}\ \bibnamefont {Thakur}}, \bibinfo
  {author} {\bibfnamefont {J.}~\bibnamefont {Gabel}}, \bibinfo {author}
  {\bibfnamefont {T.-L.}\ \bibnamefont {Lee}}, \bibinfo {author} {\bibfnamefont
  {J.}~\bibnamefont {Sch{\"a}fer}}, \bibinfo {author} {\bibfnamefont
  {S.}~\bibnamefont {Moser}}, \bibinfo {author} {\bibfnamefont
  {D.}~\bibnamefont {Di~Sante}}, \bibinfo {author} {\bibfnamefont
  {R.}~\bibnamefont {Claessen}},\ and\ \bibinfo {author} {\bibfnamefont
  {G.}~\bibnamefont {Sangiovanni}},\ }\bibfield  {title} {\bibinfo {title}
  {Design and realization of topological dirac fermions on a triangular
  lattice},\ }\href {https://doi.org/10.1038/s41467-021-25627-y} {\bibfield
  {journal} {\bibinfo  {journal} {Nature Communications}\ }\textbf {\bibinfo
  {volume} {12}},\ \bibinfo {pages} {5396} (\bibinfo {year}
  {2021})}\BibitemShut {NoStop}%
\bibitem [{\citenamefont {Schmitt}\ \emph {et~al.}(2023)\citenamefont
  {Schmitt}, \citenamefont {Erhardt}, \citenamefont {Eck}, \citenamefont
  {Schmitt}, \citenamefont {Lee}, \citenamefont {Wagner}, \citenamefont
  {Keßler}, \citenamefont {Kamp}, \citenamefont {Kim}, \citenamefont {Cacho},
  \citenamefont {Lee}, \citenamefont {Sangiovanni}, \citenamefont {Moser},\
  and\ \citenamefont {Claessen}}]{schmitt2023}%
  \BibitemOpen
  \bibfield  {author} {\bibinfo {author} {\bibfnamefont {C.}~\bibnamefont
  {Schmitt}}, \bibinfo {author} {\bibfnamefont {J.}~\bibnamefont {Erhardt}},
  \bibinfo {author} {\bibfnamefont {P.}~\bibnamefont {Eck}}, \bibinfo {author}
  {\bibfnamefont {M.}~\bibnamefont {Schmitt}}, \bibinfo {author} {\bibfnamefont
  {K.}~\bibnamefont {Lee}}, \bibinfo {author} {\bibfnamefont {T.}~\bibnamefont
  {Wagner}}, \bibinfo {author} {\bibfnamefont {P.}~\bibnamefont {Keßler}},
  \bibinfo {author} {\bibfnamefont {M.}~\bibnamefont {Kamp}}, \bibinfo {author}
  {\bibfnamefont {T.}~\bibnamefont {Kim}}, \bibinfo {author} {\bibfnamefont
  {C.}~\bibnamefont {Cacho}}, \bibinfo {author} {\bibfnamefont {T.-L.}\
  \bibnamefont {Lee}}, \bibinfo {author} {\bibfnamefont {G.}~\bibnamefont
  {Sangiovanni}}, \bibinfo {author} {\bibfnamefont {S.}~\bibnamefont {Moser}},\
  and\ \bibinfo {author} {\bibfnamefont {R.}~\bibnamefont {Claessen}},\
  }\href@noop {} {\bibinfo {title} {Stabilizing an atomically thin quantum spin
  hall insulator at ambient conditions: Graphene-intercalation of indenene}}
  (\bibinfo {year} {2023}),\ \Eprint {https://arxiv.org/abs/2305.07807}
  {arXiv:2305.07807 [cond-mat.mtrl-sci]} \BibitemShut {NoStop}%
\bibitem [{\citenamefont {Bampoulis}\ \emph {et~al.}(2023)\citenamefont
  {Bampoulis}, \citenamefont {Castenmiller}, \citenamefont {Klaassen},
  \citenamefont {van Mil}, \citenamefont {Liu}, \citenamefont {Liu},
  \citenamefont {Yao}, \citenamefont {Ezawa}, \citenamefont {Rudenko},\ and\
  \citenamefont {Zandvliet}}]{Bampoulis2023}%
  \BibitemOpen
  \bibfield  {author} {\bibinfo {author} {\bibfnamefont {P.}~\bibnamefont
  {Bampoulis}}, \bibinfo {author} {\bibfnamefont {C.}~\bibnamefont
  {Castenmiller}}, \bibinfo {author} {\bibfnamefont {D.~J.}\ \bibnamefont
  {Klaassen}}, \bibinfo {author} {\bibfnamefont {J.}~\bibnamefont {van Mil}},
  \bibinfo {author} {\bibfnamefont {Y.}~\bibnamefont {Liu}}, \bibinfo {author}
  {\bibfnamefont {C.-C.}\ \bibnamefont {Liu}}, \bibinfo {author} {\bibfnamefont
  {Y.}~\bibnamefont {Yao}}, \bibinfo {author} {\bibfnamefont {M.}~\bibnamefont
  {Ezawa}}, \bibinfo {author} {\bibfnamefont {A.~N.}\ \bibnamefont {Rudenko}},\
  and\ \bibinfo {author} {\bibfnamefont {H.~J.~W.}\ \bibnamefont {Zandvliet}},\
  }\bibfield  {title} {\bibinfo {title} {Quantum spin hall states and
  topological phase transition in germanene},\ }\href
  {https://doi.org/10.1103/PhysRevLett.130.196401} {\bibfield  {journal}
  {\bibinfo  {journal} {Phys. Rev. Lett.}\ }\textbf {\bibinfo {volume} {130}},\
  \bibinfo {pages} {196401} (\bibinfo {year} {2023})}\BibitemShut {NoStop}%
\bibitem [{\citenamefont {Tao}\ \emph {et~al.}(2023)\citenamefont {Tao},
  \citenamefont {Shen}, \citenamefont {Zhao}, \citenamefont {Hu}, \citenamefont
  {Li}, \citenamefont {Jiang}, \citenamefont {Li}, \citenamefont {Watanabe},
  \citenamefont {Taniguchi}, \citenamefont {MacDonald}, \citenamefont {Shan},\
  and\ \citenamefont {Mak}}]{tao2023giant}%
  \BibitemOpen
  \bibfield  {author} {\bibinfo {author} {\bibfnamefont {Z.}~\bibnamefont
  {Tao}}, \bibinfo {author} {\bibfnamefont {B.}~\bibnamefont {Shen}}, \bibinfo
  {author} {\bibfnamefont {W.}~\bibnamefont {Zhao}}, \bibinfo {author}
  {\bibfnamefont {N.~C.}\ \bibnamefont {Hu}}, \bibinfo {author} {\bibfnamefont
  {T.}~\bibnamefont {Li}}, \bibinfo {author} {\bibfnamefont {S.}~\bibnamefont
  {Jiang}}, \bibinfo {author} {\bibfnamefont {L.}~\bibnamefont {Li}}, \bibinfo
  {author} {\bibfnamefont {K.}~\bibnamefont {Watanabe}}, \bibinfo {author}
  {\bibfnamefont {T.}~\bibnamefont {Taniguchi}}, \bibinfo {author}
  {\bibfnamefont {A.~H.}\ \bibnamefont {MacDonald}}, \bibinfo {author}
  {\bibfnamefont {J.}~\bibnamefont {Shan}},\ and\ \bibinfo {author}
  {\bibfnamefont {K.~F.}\ \bibnamefont {Mak}},\ }\href@noop {} {\bibinfo
  {title} {Giant spin hall effect in ab-stacked mote2/wse2 bilayers}} (\bibinfo
  {year} {2023}),\ \Eprint {https://arxiv.org/abs/2303.12881} {arXiv:2303.12881
  [cond-mat.mes-hall]} \BibitemShut {NoStop}%
\bibitem [{\citenamefont {Liu}\ \emph {et~al.}(2012{\natexlab{a}})\citenamefont
  {Liu}, \citenamefont {Pai}, \citenamefont {Li}, \citenamefont {Tseng},
  \citenamefont {Ralph},\ and\ \citenamefont {Buhrman}}]{Luqiao2012}%
  \BibitemOpen
  \bibfield  {author} {\bibinfo {author} {\bibfnamefont {L.}~\bibnamefont
  {Liu}}, \bibinfo {author} {\bibfnamefont {C.-F.}\ \bibnamefont {Pai}},
  \bibinfo {author} {\bibfnamefont {Y.}~\bibnamefont {Li}}, \bibinfo {author}
  {\bibfnamefont {H.~W.}\ \bibnamefont {Tseng}}, \bibinfo {author}
  {\bibfnamefont {D.~C.}\ \bibnamefont {Ralph}},\ and\ \bibinfo {author}
  {\bibfnamefont {R.~A.}\ \bibnamefont {Buhrman}},\ }\bibfield  {title}
  {\bibinfo {title} {Spin-torque switching with the giant spin hall effect of
  tantalum},\ }\href {https://doi.org/10.1126/science.1218197} {\bibfield
  {journal} {\bibinfo  {journal} {Science}\ }\textbf {\bibinfo {volume}
  {336}},\ \bibinfo {pages} {555} (\bibinfo {year} {2012}{\natexlab{a}})},\
  \Eprint
  {https://arxiv.org/abs/https://www.science.org/doi/pdf/10.1126/science.1218197}
  {https://www.science.org/doi/pdf/10.1126/science.1218197} \BibitemShut
  {NoStop}%
\bibitem [{\citenamefont {Liu}\ \emph {et~al.}(2012{\natexlab{b}})\citenamefont
  {Liu}, \citenamefont {Lee}, \citenamefont {Gudmundsen}, \citenamefont
  {Ralph},\ and\ \citenamefont {Buhrman}}]{Liu2012}%
  \BibitemOpen
  \bibfield  {author} {\bibinfo {author} {\bibfnamefont {L.}~\bibnamefont
  {Liu}}, \bibinfo {author} {\bibfnamefont {O.~J.}\ \bibnamefont {Lee}},
  \bibinfo {author} {\bibfnamefont {T.~J.}\ \bibnamefont {Gudmundsen}},
  \bibinfo {author} {\bibfnamefont {D.~C.}\ \bibnamefont {Ralph}},\ and\
  \bibinfo {author} {\bibfnamefont {R.~A.}\ \bibnamefont {Buhrman}},\
  }\bibfield  {title} {\bibinfo {title} {Current-induced switching of
  perpendicularly magnetized magnetic layers using spin torque from the spin
  hall effect},\ }\href {https://doi.org/10.1103/PhysRevLett.109.096602}
  {\bibfield  {journal} {\bibinfo  {journal} {Phys. Rev. Lett.}\ }\textbf
  {\bibinfo {volume} {109}},\ \bibinfo {pages} {096602} (\bibinfo {year}
  {2012}{\natexlab{b}})}\BibitemShut {NoStop}%
\bibitem [{\citenamefont {\ifmmode \check{Z}\else
  \v{Z}\fi{}uti\ifmmode~\acute{c}\else \'{c}\fi{}}\ \emph
  {et~al.}(2004)\citenamefont {\ifmmode \check{Z}\else
  \v{Z}\fi{}uti\ifmmode~\acute{c}\else \'{c}\fi{}}, \citenamefont {Fabian},\
  and\ \citenamefont {Das~Sarma}}]{zutic2004}%
  \BibitemOpen
  \bibfield  {author} {\bibinfo {author} {\bibfnamefont {I.}~\bibnamefont
  {\ifmmode \check{Z}\else \v{Z}\fi{}uti\ifmmode~\acute{c}\else \'{c}\fi{}}},
  \bibinfo {author} {\bibfnamefont {J.}~\bibnamefont {Fabian}},\ and\ \bibinfo
  {author} {\bibfnamefont {S.}~\bibnamefont {Das~Sarma}},\ }\bibfield  {title}
  {\bibinfo {title} {Spintronics: Fundamentals and applications},\ }\href
  {https://doi.org/10.1103/RevModPhys.76.323} {\bibfield  {journal} {\bibinfo
  {journal} {Rev. Mod. Phys.}\ }\textbf {\bibinfo {volume} {76}},\ \bibinfo
  {pages} {323} (\bibinfo {year} {2004})}\BibitemShut {NoStop}%
\bibitem [{\citenamefont {Sinova}\ \emph {et~al.}(2015)\citenamefont {Sinova},
  \citenamefont {Valenzuela}, \citenamefont {Wunderlich}, \citenamefont
  {Back},\ and\ \citenamefont {Jungwirth}}]{sinova2015}%
  \BibitemOpen
  \bibfield  {author} {\bibinfo {author} {\bibfnamefont {J.}~\bibnamefont
  {Sinova}}, \bibinfo {author} {\bibfnamefont {S.~O.}\ \bibnamefont
  {Valenzuela}}, \bibinfo {author} {\bibfnamefont {J.}~\bibnamefont
  {Wunderlich}}, \bibinfo {author} {\bibfnamefont {C.~H.}\ \bibnamefont
  {Back}},\ and\ \bibinfo {author} {\bibfnamefont {T.}~\bibnamefont
  {Jungwirth}},\ }\bibfield  {title} {\bibinfo {title} {Spin hall effects},\
  }\href {https://doi.org/10.1103/RevModPhys.87.1213} {\bibfield  {journal}
  {\bibinfo  {journal} {Rev. Mod. Phys.}\ }\textbf {\bibinfo {volume} {87}},\
  \bibinfo {pages} {1213} (\bibinfo {year} {2015})}\BibitemShut {NoStop}%
\bibitem [{\citenamefont {Lee}(2011)}]{Lee2011}%
  \BibitemOpen
  \bibfield  {author} {\bibinfo {author} {\bibfnamefont {D.-H.}\ \bibnamefont
  {Lee}},\ }\bibfield  {title} {\bibinfo {title} {Effects of interaction on
  quantum spin hall insulators},\ }\href
  {https://doi.org/10.1103/PhysRevLett.107.166806} {\bibfield  {journal}
  {\bibinfo  {journal} {Phys. Rev. Lett.}\ }\textbf {\bibinfo {volume} {107}},\
  \bibinfo {pages} {166806} (\bibinfo {year} {2011})}\BibitemShut {NoStop}%
\bibitem [{\citenamefont {Griset}\ and\ \citenamefont {Xu}(2012)}]{griset2012}%
  \BibitemOpen
  \bibfield  {author} {\bibinfo {author} {\bibfnamefont {C.}~\bibnamefont
  {Griset}}\ and\ \bibinfo {author} {\bibfnamefont {C.}~\bibnamefont {Xu}},\
  }\bibfield  {title} {\bibinfo {title} {Phase diagram of the kane-mele-hubbard
  model},\ }\href {https://doi.org/10.1103/PhysRevB.85.045123} {\bibfield
  {journal} {\bibinfo  {journal} {Phys. Rev. B}\ }\textbf {\bibinfo {volume}
  {85}},\ \bibinfo {pages} {045123} (\bibinfo {year} {2012})}\BibitemShut
  {NoStop}%
\bibitem [{\citenamefont {Hutchinson}\ \emph {et~al.}(2021)\citenamefont
  {Hutchinson}, \citenamefont {Klein},\ and\ \citenamefont
  {Le~Hur}}]{Hutchinson2021}%
  \BibitemOpen
  \bibfield  {author} {\bibinfo {author} {\bibfnamefont {J.}~\bibnamefont
  {Hutchinson}}, \bibinfo {author} {\bibfnamefont {P.~W.}\ \bibnamefont
  {Klein}},\ and\ \bibinfo {author} {\bibfnamefont {K.}~\bibnamefont
  {Le~Hur}},\ }\bibfield  {title} {\bibinfo {title} {Analytical approach for
  the mott transition in the kane-mele-hubbard model},\ }\href
  {https://doi.org/10.1103/PhysRevB.104.075120} {\bibfield  {journal} {\bibinfo
   {journal} {Phys. Rev. B}\ }\textbf {\bibinfo {volume} {104}},\ \bibinfo
  {pages} {075120} (\bibinfo {year} {2021})}\BibitemShut {NoStop}%
\bibitem [{\citenamefont {Hohenadler}\ \emph {et~al.}(2011)\citenamefont
  {Hohenadler}, \citenamefont {Lang},\ and\ \citenamefont
  {Assaad}}]{Hohenadler2011}%
  \BibitemOpen
  \bibfield  {author} {\bibinfo {author} {\bibfnamefont {M.}~\bibnamefont
  {Hohenadler}}, \bibinfo {author} {\bibfnamefont {T.~C.}\ \bibnamefont
  {Lang}},\ and\ \bibinfo {author} {\bibfnamefont {F.~F.}\ \bibnamefont
  {Assaad}},\ }\bibfield  {title} {\bibinfo {title} {Correlation effects in
  quantum spin-hall insulators: A quantum monte carlo study},\ }\href
  {https://doi.org/10.1103/PhysRevLett.106.100403} {\bibfield  {journal}
  {\bibinfo  {journal} {Phys. Rev. Lett.}\ }\textbf {\bibinfo {volume} {106}},\
  \bibinfo {pages} {100403} (\bibinfo {year} {2011})}\BibitemShut {NoStop}%
\bibitem [{\citenamefont {Hohenadler}\ \emph {et~al.}(2012)\citenamefont
  {Hohenadler}, \citenamefont {Meng}, \citenamefont {Lang}, \citenamefont
  {Wessel}, \citenamefont {Muramatsu},\ and\ \citenamefont
  {Assaad}}]{Hohenadler2012}%
  \BibitemOpen
  \bibfield  {author} {\bibinfo {author} {\bibfnamefont {M.}~\bibnamefont
  {Hohenadler}}, \bibinfo {author} {\bibfnamefont {Z.~Y.}\ \bibnamefont
  {Meng}}, \bibinfo {author} {\bibfnamefont {T.~C.}\ \bibnamefont {Lang}},
  \bibinfo {author} {\bibfnamefont {S.}~\bibnamefont {Wessel}}, \bibinfo
  {author} {\bibfnamefont {A.}~\bibnamefont {Muramatsu}},\ and\ \bibinfo
  {author} {\bibfnamefont {F.~F.}\ \bibnamefont {Assaad}},\ }\bibfield  {title}
  {\bibinfo {title} {Quantum phase transitions in the kane-mele-hubbard
  model},\ }\href {https://doi.org/10.1103/PhysRevB.85.115132} {\bibfield
  {journal} {\bibinfo  {journal} {Phys. Rev. B}\ }\textbf {\bibinfo {volume}
  {85}},\ \bibinfo {pages} {115132} (\bibinfo {year} {2012})}\BibitemShut
  {NoStop}%
\bibitem [{\citenamefont {Rachel}\ and\ \citenamefont
  {Le~Hur}(2010)}]{Rachel2010}%
  \BibitemOpen
  \bibfield  {author} {\bibinfo {author} {\bibfnamefont {S.}~\bibnamefont
  {Rachel}}\ and\ \bibinfo {author} {\bibfnamefont {K.}~\bibnamefont
  {Le~Hur}},\ }\bibfield  {title} {\bibinfo {title} {Topological insulators and
  mott physics from the hubbard interaction},\ }\href
  {https://doi.org/10.1103/PhysRevB.82.075106} {\bibfield  {journal} {\bibinfo
  {journal} {Phys. Rev. B}\ }\textbf {\bibinfo {volume} {82}},\ \bibinfo
  {pages} {075106} (\bibinfo {year} {2010})}\BibitemShut {NoStop}%
\bibitem [{\citenamefont {Reuther}\ \emph {et~al.}(2012)\citenamefont
  {Reuther}, \citenamefont {Thomale},\ and\ \citenamefont
  {Rachel}}]{Reuther2012}%
  \BibitemOpen
  \bibfield  {author} {\bibinfo {author} {\bibfnamefont {J.}~\bibnamefont
  {Reuther}}, \bibinfo {author} {\bibfnamefont {R.}~\bibnamefont {Thomale}},\
  and\ \bibinfo {author} {\bibfnamefont {S.}~\bibnamefont {Rachel}},\
  }\bibfield  {title} {\bibinfo {title} {Magnetic ordering phenomena of
  interacting quantum spin hall models},\ }\href
  {https://doi.org/10.1103/PhysRevB.86.155127} {\bibfield  {journal} {\bibinfo
  {journal} {Phys. Rev. B}\ }\textbf {\bibinfo {volume} {86}},\ \bibinfo
  {pages} {155127} (\bibinfo {year} {2012})}\BibitemShut {NoStop}%
\bibitem [{\citenamefont {Yu}\ \emph {et~al.}(2011)\citenamefont {Yu},
  \citenamefont {Xie},\ and\ \citenamefont {Li}}]{Yu2011}%
  \BibitemOpen
  \bibfield  {author} {\bibinfo {author} {\bibfnamefont {S.-L.}\ \bibnamefont
  {Yu}}, \bibinfo {author} {\bibfnamefont {X.~C.}\ \bibnamefont {Xie}},\ and\
  \bibinfo {author} {\bibfnamefont {J.-X.}\ \bibnamefont {Li}},\ }\bibfield
  {title} {\bibinfo {title} {Mott physics and topological phase transition in
  correlated dirac fermions},\ }\href
  {https://doi.org/10.1103/PhysRevLett.107.010401} {\bibfield  {journal}
  {\bibinfo  {journal} {Phys. Rev. Lett.}\ }\textbf {\bibinfo {volume} {107}},\
  \bibinfo {pages} {010401} (\bibinfo {year} {2011})}\BibitemShut {NoStop}%
\bibitem [{\citenamefont {Zheng}\ \emph {et~al.}(2011)\citenamefont {Zheng},
  \citenamefont {Zhang},\ and\ \citenamefont {Wu}}]{Zheng2011}%
  \BibitemOpen
  \bibfield  {author} {\bibinfo {author} {\bibfnamefont {D.}~\bibnamefont
  {Zheng}}, \bibinfo {author} {\bibfnamefont {G.-M.}\ \bibnamefont {Zhang}},\
  and\ \bibinfo {author} {\bibfnamefont {C.}~\bibnamefont {Wu}},\ }\bibfield
  {title} {\bibinfo {title} {Particle-hole symmetry and interaction effects in
  the kane-mele-hubbard model},\ }\href
  {https://doi.org/10.1103/PhysRevB.84.205121} {\bibfield  {journal} {\bibinfo
  {journal} {Phys. Rev. B}\ }\textbf {\bibinfo {volume} {84}},\ \bibinfo
  {pages} {205121} (\bibinfo {year} {2011})}\BibitemShut {NoStop}%
\bibitem [{\citenamefont {Wu}\ \emph {et~al.}(2012)\citenamefont {Wu},
  \citenamefont {Rachel}, \citenamefont {Liu},\ and\ \citenamefont
  {Le~Hur}}]{Wu2012}%
  \BibitemOpen
  \bibfield  {author} {\bibinfo {author} {\bibfnamefont {W.}~\bibnamefont
  {Wu}}, \bibinfo {author} {\bibfnamefont {S.}~\bibnamefont {Rachel}}, \bibinfo
  {author} {\bibfnamefont {W.-M.}\ \bibnamefont {Liu}},\ and\ \bibinfo {author}
  {\bibfnamefont {K.}~\bibnamefont {Le~Hur}},\ }\bibfield  {title} {\bibinfo
  {title} {Quantum spin hall insulators with interactions and lattice
  anisotropy},\ }\href {https://doi.org/10.1103/PhysRevB.85.205102} {\bibfield
  {journal} {\bibinfo  {journal} {Phys. Rev. B}\ }\textbf {\bibinfo {volume}
  {85}},\ \bibinfo {pages} {205102} (\bibinfo {year} {2012})}\BibitemShut
  {NoStop}%
\bibitem [{\citenamefont {Laubach}\ \emph {et~al.}(2014)\citenamefont
  {Laubach}, \citenamefont {Reuther}, \citenamefont {Thomale},\ and\
  \citenamefont {Rachel}}]{Laubach2014}%
  \BibitemOpen
  \bibfield  {author} {\bibinfo {author} {\bibfnamefont {M.}~\bibnamefont
  {Laubach}}, \bibinfo {author} {\bibfnamefont {J.}~\bibnamefont {Reuther}},
  \bibinfo {author} {\bibfnamefont {R.}~\bibnamefont {Thomale}},\ and\ \bibinfo
  {author} {\bibfnamefont {S.}~\bibnamefont {Rachel}},\ }\bibfield  {title}
  {\bibinfo {title} {Rashba spin-orbit coupling in the kane-mele-hubbard
  model},\ }\href {https://doi.org/10.1103/PhysRevB.90.165136} {\bibfield
  {journal} {\bibinfo  {journal} {Phys. Rev. B}\ }\textbf {\bibinfo {volume}
  {90}},\ \bibinfo {pages} {165136} (\bibinfo {year} {2014})}\BibitemShut
  {NoStop}%
\bibitem [{\citenamefont {Hung}\ \emph {et~al.}(2013)\citenamefont {Hung},
  \citenamefont {Wang}, \citenamefont {Gu},\ and\ \citenamefont
  {Fiete}}]{Hung2013}%
  \BibitemOpen
  \bibfield  {author} {\bibinfo {author} {\bibfnamefont {H.-H.}\ \bibnamefont
  {Hung}}, \bibinfo {author} {\bibfnamefont {L.}~\bibnamefont {Wang}}, \bibinfo
  {author} {\bibfnamefont {Z.-C.}\ \bibnamefont {Gu}},\ and\ \bibinfo {author}
  {\bibfnamefont {G.~A.}\ \bibnamefont {Fiete}},\ }\bibfield  {title} {\bibinfo
  {title} {Topological phase transition in a generalized kane-mele-hubbard
  model: A combined quantum monte carlo and green's function study},\ }\href
  {https://doi.org/10.1103/PhysRevB.87.121113} {\bibfield  {journal} {\bibinfo
  {journal} {Phys. Rev. B}\ }\textbf {\bibinfo {volume} {87}},\ \bibinfo
  {pages} {121113} (\bibinfo {year} {2013})}\BibitemShut {NoStop}%
\bibitem [{\citenamefont {Richter}\ \emph {et~al.}(2021)\citenamefont
  {Richter}, \citenamefont {Graspeuntner}, \citenamefont {Sch\"afer},
  \citenamefont {Wentzell},\ and\ \citenamefont {Aichhorn}}]{Richter2021}%
  \BibitemOpen
  \bibfield  {author} {\bibinfo {author} {\bibfnamefont {M.}~\bibnamefont
  {Richter}}, \bibinfo {author} {\bibfnamefont {J.}~\bibnamefont
  {Graspeuntner}}, \bibinfo {author} {\bibfnamefont {T.}~\bibnamefont
  {Sch\"afer}}, \bibinfo {author} {\bibfnamefont {N.}~\bibnamefont
  {Wentzell}},\ and\ \bibinfo {author} {\bibfnamefont {M.}~\bibnamefont
  {Aichhorn}},\ }\bibfield  {title} {\bibinfo {title} {Comparing the effective
  enhancement of local and nonlocal spin-orbit couplings on honeycomb lattices
  due to strong electronic correlations},\ }\href
  {https://doi.org/10.1103/PhysRevB.104.195107} {\bibfield  {journal} {\bibinfo
   {journal} {Phys. Rev. B}\ }\textbf {\bibinfo {volume} {104}},\ \bibinfo
  {pages} {195107} (\bibinfo {year} {2021})}\BibitemShut {NoStop}%
\bibitem [{\citenamefont {Mai}\ \emph {et~al.}(2023)\citenamefont {Mai},
  \citenamefont {Zhao}, \citenamefont {Feldman},\ and\ \citenamefont
  {Phillips}}]{mai2023}%
  \BibitemOpen
  \bibfield  {author} {\bibinfo {author} {\bibfnamefont {P.}~\bibnamefont
  {Mai}}, \bibinfo {author} {\bibfnamefont {J.}~\bibnamefont {Zhao}}, \bibinfo
  {author} {\bibfnamefont {B.~E.}\ \bibnamefont {Feldman}},\ and\ \bibinfo
  {author} {\bibfnamefont {P.~W.}\ \bibnamefont {Phillips}},\ }\href@noop {}
  {\bibinfo {title} {1/4 is the new 1/2: Interaction-induced unification of
  quantum anomalous and spin hall effects}} (\bibinfo {year} {2023}),\ \Eprint
  {https://arxiv.org/abs/2210.11486} {arXiv:2210.11486 [cond-mat.mes-hall]}
  \BibitemShut {NoStop}%
\bibitem [{\citenamefont {Hohenadler}\ and\ \citenamefont
  {Assaad}(2013)}]{rev_Hohenadler_2013}%
  \BibitemOpen
  \bibfield  {author} {\bibinfo {author} {\bibfnamefont {M.}~\bibnamefont
  {Hohenadler}}\ and\ \bibinfo {author} {\bibfnamefont {F.~F.}\ \bibnamefont
  {Assaad}},\ }\bibfield  {title} {\bibinfo {title} {Correlation effects in
  two-dimensional topological insulators},\ }\href
  {https://doi.org/10.1088/0953-8984/25/14/143201} {\bibfield  {journal}
  {\bibinfo  {journal} {Journal of Physics: Condensed Matter}\ }\textbf
  {\bibinfo {volume} {25}},\ \bibinfo {pages} {143201} (\bibinfo {year}
  {2013})}\BibitemShut {NoStop}%
\bibitem [{\citenamefont {Rachel}(2018)}]{Rachel_2018}%
  \BibitemOpen
  \bibfield  {author} {\bibinfo {author} {\bibfnamefont {S.}~\bibnamefont
  {Rachel}},\ }\bibfield  {title} {\bibinfo {title} {Interacting topological
  insulators: a review},\ }\href {https://doi.org/10.1088/1361-6633/aad6a6}
  {\bibfield  {journal} {\bibinfo  {journal} {Reports on Progress in Physics}\
  }\textbf {\bibinfo {volume} {81}},\ \bibinfo {pages} {116501} (\bibinfo
  {year} {2018})}\BibitemShut {NoStop}%
\bibitem [{\citenamefont {MENG}\ \emph {et~al.}(2014)\citenamefont {MENG},
  \citenamefont {HUNG},\ and\ \citenamefont {LANG}}]{Meng2014}%
  \BibitemOpen
  \bibfield  {author} {\bibinfo {author} {\bibfnamefont {Z.~Y.}\ \bibnamefont
  {MENG}}, \bibinfo {author} {\bibfnamefont {H.-H.}\ \bibnamefont {HUNG}},\
  and\ \bibinfo {author} {\bibfnamefont {T.~C.}\ \bibnamefont {LANG}},\
  }\bibfield  {title} {\bibinfo {title} {The characterization of topological
  properties in quantum monte carlo simulations of the kane–mele–hubbard
  model},\ }\href {https://doi.org/10.1142/S0217984914300014} {\bibfield
  {journal} {\bibinfo  {journal} {Modern Physics Letters B}\ }\textbf {\bibinfo
  {volume} {28}},\ \bibinfo {pages} {1430001} (\bibinfo {year} {2014})},\
  \Eprint {https://arxiv.org/abs/https://doi.org/10.1142/S0217984914300014}
  {https://doi.org/10.1142/S0217984914300014} \BibitemShut {NoStop}%
\bibitem [{\citenamefont {Sorella}\ \emph {et~al.}(2012)\citenamefont
  {Sorella}, \citenamefont {Otsuka},\ and\ \citenamefont
  {Yunoki}}]{Sorella2012}%
  \BibitemOpen
  \bibfield  {author} {\bibinfo {author} {\bibfnamefont {S.}~\bibnamefont
  {Sorella}}, \bibinfo {author} {\bibfnamefont {Y.}~\bibnamefont {Otsuka}},\
  and\ \bibinfo {author} {\bibfnamefont {S.}~\bibnamefont {Yunoki}},\
  }\bibfield  {title} {\bibinfo {title} {Absence of a spin liquid phase in the
  hubbard model on the honeycomb lattice},\ }\href
  {https://doi.org/10.1038/srep00992} {\bibfield  {journal} {\bibinfo
  {journal} {Scientific Reports}\ }\textbf {\bibinfo {volume} {2}},\ \bibinfo
  {pages} {992} (\bibinfo {year} {2012})}\BibitemShut {NoStop}%
\bibitem [{\citenamefont {Yamaji}\ and\ \citenamefont
  {Imada}(2011)}]{Yamaji2011}%
  \BibitemOpen
  \bibfield  {author} {\bibinfo {author} {\bibfnamefont {Y.}~\bibnamefont
  {Yamaji}}\ and\ \bibinfo {author} {\bibfnamefont {M.}~\bibnamefont {Imada}},\
  }\bibfield  {title} {\bibinfo {title} {Mott physics on helical edges of
  two-dimensional topological insulators},\ }\href
  {https://doi.org/10.1103/PhysRevB.83.205122} {\bibfield  {journal} {\bibinfo
  {journal} {Phys. Rev. B}\ }\textbf {\bibinfo {volume} {83}},\ \bibinfo
  {pages} {205122} (\bibinfo {year} {2011})}\BibitemShut {NoStop}%
\bibitem [{\citenamefont {Yoshida}\ and\ \citenamefont
  {Kawakami}(2016)}]{Yoshida2016}%
  \BibitemOpen
  \bibfield  {author} {\bibinfo {author} {\bibfnamefont {T.}~\bibnamefont
  {Yoshida}}\ and\ \bibinfo {author} {\bibfnamefont {N.}~\bibnamefont
  {Kawakami}},\ }\bibfield  {title} {\bibinfo {title} {Topological edge mott
  insulating state in two dimensions at finite temperatures: Bulk and edge
  analysis},\ }\href {https://doi.org/10.1103/PhysRevB.94.085149} {\bibfield
  {journal} {\bibinfo  {journal} {Phys. Rev. B}\ }\textbf {\bibinfo {volume}
  {94}},\ \bibinfo {pages} {085149} (\bibinfo {year} {2016})}\BibitemShut
  {NoStop}%
\bibitem [{\citenamefont {Wagner}\ \emph {et~al.}(2023)\citenamefont {Wagner},
  \citenamefont {Crippa}, \citenamefont {Amaricci}, \citenamefont {Hansmann},
  \citenamefont {Klett}, \citenamefont {König}, \citenamefont {Schäfer},
  \citenamefont {Sante}, \citenamefont {Cano}, \citenamefont {Millis},
  \citenamefont {Georges},\ and\ \citenamefont {Sangiovanni}}]{wagner2023mott}%
  \BibitemOpen
  \bibfield  {author} {\bibinfo {author} {\bibfnamefont {N.}~\bibnamefont
  {Wagner}}, \bibinfo {author} {\bibfnamefont {L.}~\bibnamefont {Crippa}},
  \bibinfo {author} {\bibfnamefont {A.}~\bibnamefont {Amaricci}}, \bibinfo
  {author} {\bibfnamefont {P.}~\bibnamefont {Hansmann}}, \bibinfo {author}
  {\bibfnamefont {M.}~\bibnamefont {Klett}}, \bibinfo {author} {\bibfnamefont
  {E.}~\bibnamefont {König}}, \bibinfo {author} {\bibfnamefont
  {T.}~\bibnamefont {Schäfer}}, \bibinfo {author} {\bibfnamefont {D.~D.}\
  \bibnamefont {Sante}}, \bibinfo {author} {\bibfnamefont {J.}~\bibnamefont
  {Cano}}, \bibinfo {author} {\bibfnamefont {A.}~\bibnamefont {Millis}},
  \bibinfo {author} {\bibfnamefont {A.}~\bibnamefont {Georges}},\ and\ \bibinfo
  {author} {\bibfnamefont {G.}~\bibnamefont {Sangiovanni}},\ }\href@noop {}
  {\bibinfo {title} {Mott insulators with boundary zeros}} (\bibinfo {year}
  {2023}),\ \Eprint {https://arxiv.org/abs/2301.05588} {arXiv:2301.05588
  [cond-mat.str-el]} \BibitemShut {NoStop}%
\bibitem [{\citenamefont {Yoshida}\ \emph {et~al.}(2012)\citenamefont
  {Yoshida}, \citenamefont {Fujimoto},\ and\ \citenamefont
  {Kawakami}}]{Yoshida2012}%
  \BibitemOpen
  \bibfield  {author} {\bibinfo {author} {\bibfnamefont {T.}~\bibnamefont
  {Yoshida}}, \bibinfo {author} {\bibfnamefont {S.}~\bibnamefont {Fujimoto}},\
  and\ \bibinfo {author} {\bibfnamefont {N.}~\bibnamefont {Kawakami}},\
  }\bibfield  {title} {\bibinfo {title} {Correlation effects on a topological
  insulator at finite temperatures},\ }\href
  {https://doi.org/10.1103/PhysRevB.85.125113} {\bibfield  {journal} {\bibinfo
  {journal} {Phys. Rev. B}\ }\textbf {\bibinfo {volume} {85}},\ \bibinfo
  {pages} {125113} (\bibinfo {year} {2012})}\BibitemShut {NoStop}%
\bibitem [{\citenamefont {Vilk}\ \emph {et~al.}(1994)\citenamefont {Vilk},
  \citenamefont {Chen},\ and\ \citenamefont {Tremblay}}]{vilk1994}%
  \BibitemOpen
  \bibfield  {author} {\bibinfo {author} {\bibfnamefont {Y.~M.}\ \bibnamefont
  {Vilk}}, \bibinfo {author} {\bibfnamefont {L.}~\bibnamefont {Chen}},\ and\
  \bibinfo {author} {\bibfnamefont {A.-M.~S.}\ \bibnamefont {Tremblay}},\
  }\bibfield  {title} {\bibinfo {title} {Theory of spin and charge fluctuations
  in the hubbard model},\ }\href {https://doi.org/10.1103/PhysRevB.49.13267}
  {\bibfield  {journal} {\bibinfo  {journal} {Phys. Rev. B}\ }\textbf {\bibinfo
  {volume} {49}},\ \bibinfo {pages} {13267} (\bibinfo {year}
  {1994})}\BibitemShut {NoStop}%
\bibitem [{\citenamefont {Vilk}\ and\ \citenamefont
  {Tremblay}(1996)}]{vilk1996}%
  \BibitemOpen
  \bibfield  {author} {\bibinfo {author} {\bibfnamefont {Y.~M.}\ \bibnamefont
  {Vilk}}\ and\ \bibinfo {author} {\bibfnamefont {A.-M.~S.}\ \bibnamefont
  {Tremblay}},\ }\bibfield  {title} {\bibinfo {title} {Destruction of
  fermi-liquid quasiparticles in two dimensions by critical fluctuations},\
  }\href {https://doi.org/10.1209/epl/i1996-00315-2} {\bibfield  {journal}
  {\bibinfo  {journal} {Europhysics Letters}\ }\textbf {\bibinfo {volume}
  {33}},\ \bibinfo {pages} {159} (\bibinfo {year} {1996})}\BibitemShut
  {NoStop}%
\bibitem [{\citenamefont {{Y.M. Vilk}}\ and\ \citenamefont {{A.-M.S.
  Tremblay}}(1997)}]{vilk1997}%
  \BibitemOpen
  \bibfield  {author} {\bibinfo {author} {\bibnamefont {{Y.M. Vilk}}}\ and\
  \bibinfo {author} {\bibnamefont {{A.-M.S. Tremblay}}},\ }\bibfield  {title}
  {\bibinfo {title} {Non-perturbative many-body approach to the hubbard model
  and single-particle pseudogap},\ }\href {https://doi.org/10.1051/jp1:1997135}
  {\bibfield  {journal} {\bibinfo  {journal} {J. Phys. I France}\ }\textbf
  {\bibinfo {volume} {7}},\ \bibinfo {pages} {1309} (\bibinfo {year}
  {1997})}\BibitemShut {NoStop}%
\bibitem [{\citenamefont {{D. Lessnich et al.}}(2024)}]{lessnichtbp}%
  \BibitemOpen
  \bibfield  {author} {\bibinfo {author} {\bibnamefont {{D. Lessnich et
  al.}}},\ }\bibfield  {title} {\bibinfo {title} {Interplay of spin-orbit
  coupling and electronic correlation effects with the two-particle
  self-consistent approach},\ }\href@noop {} {\bibfield  {journal} {\bibinfo
  {journal} {in preparation}\ } (\bibinfo {year} {2024})}\BibitemShut {NoStop}%
\bibitem [{\citenamefont {Martin}\ \emph {et~al.}(2023)\citenamefont {Martin},
  \citenamefont {Gauvin-Ndiaye},\ and\ \citenamefont {Tremblay}}]{martin2023}%
  \BibitemOpen
  \bibfield  {author} {\bibinfo {author} {\bibfnamefont {N.}~\bibnamefont
  {Martin}}, \bibinfo {author} {\bibfnamefont {C.}~\bibnamefont
  {Gauvin-Ndiaye}},\ and\ \bibinfo {author} {\bibfnamefont {A.-M.~S.}\
  \bibnamefont {Tremblay}},\ }\bibfield  {title} {\bibinfo {title} {Nonlocal
  corrections to dynamical mean-field theory from the two-particle
  self-consistent method},\ }\href
  {https://doi.org/10.1103/PhysRevB.107.075158} {\bibfield  {journal} {\bibinfo
   {journal} {Phys. Rev. B}\ }\textbf {\bibinfo {volume} {107}},\ \bibinfo
  {pages} {075158} (\bibinfo {year} {2023})}\BibitemShut {NoStop}%
\bibitem [{\citenamefont {Zantout}\ \emph {et~al.}(2023)\citenamefont
  {Zantout}, \citenamefont {Backes}, \citenamefont {Razpopov}, \citenamefont
  {Lessnich},\ and\ \citenamefont {Valent\'{\i}}}]{zantout2023}%
  \BibitemOpen
  \bibfield  {author} {\bibinfo {author} {\bibfnamefont {K.}~\bibnamefont
  {Zantout}}, \bibinfo {author} {\bibfnamefont {S.}~\bibnamefont {Backes}},
  \bibinfo {author} {\bibfnamefont {A.}~\bibnamefont {Razpopov}}, \bibinfo
  {author} {\bibfnamefont {D.}~\bibnamefont {Lessnich}},\ and\ \bibinfo
  {author} {\bibfnamefont {R.}~\bibnamefont {Valent\'{\i}}},\ }\bibfield
  {title} {\bibinfo {title} {Improved effective vertices in the multiorbital
  two-particle self-consistent method from dynamical mean-field theory},\
  }\href {https://doi.org/10.1103/PhysRevB.107.235101} {\bibfield  {journal}
  {\bibinfo  {journal} {Phys. Rev. B}\ }\textbf {\bibinfo {volume} {107}},\
  \bibinfo {pages} {235101} (\bibinfo {year} {2023})}\BibitemShut {NoStop}%
\bibitem [{\citenamefont {Simard}\ and\ \citenamefont
  {Werner}(2023)}]{simard2023}%
  \BibitemOpen
  \bibfield  {author} {\bibinfo {author} {\bibfnamefont {O.}~\bibnamefont
  {Simard}}\ and\ \bibinfo {author} {\bibfnamefont {P.}~\bibnamefont
  {Werner}},\ }\href@noop {} {\bibinfo {title} {Dynamical mean field theory
  extension to the nonequilibrium two-particle self-consistent approach}}
  (\bibinfo {year} {2023}),\ \Eprint {https://arxiv.org/abs/2302.14134}
  {arXiv:2302.14134 [cond-mat.str-el]} \BibitemShut {NoStop}%
\bibitem [{\citenamefont {Schäfer}\ \emph {et~al.}(2021)\citenamefont
  {Schäfer}, \citenamefont {Wentzell}, \citenamefont {Šimkovic},
  \citenamefont {He}, \citenamefont {Hille}, \citenamefont {Klett},
  \citenamefont {Eckhardt}, \citenamefont {Arzhang}, \citenamefont {Harkov},
  \citenamefont {Le~Régent}, \citenamefont {Kirsch}, \citenamefont {Wang},
  \citenamefont {Kim}, \citenamefont {Kozik}, \citenamefont {Stepanov},
  \citenamefont {Kauch}, \citenamefont {Andergassen}, \citenamefont {Hansmann},
  \citenamefont {Rohe}, \citenamefont {Vilk}, \citenamefont {LeBlanc},
  \citenamefont {Zhang}, \citenamefont {Tremblay}, \citenamefont {Ferrero},
  \citenamefont {Parcollet},\ and\ \citenamefont {Georges}}]{Schaefer2021}%
  \BibitemOpen
  \bibfield  {author} {\bibinfo {author} {\bibfnamefont {T.}~\bibnamefont
  {Schäfer}}, \bibinfo {author} {\bibfnamefont {N.}~\bibnamefont {Wentzell}},
  \bibinfo {author} {\bibfnamefont {F.}~\bibnamefont {Šimkovic}}, \bibinfo
  {author} {\bibfnamefont {Y.-Y.}\ \bibnamefont {He}}, \bibinfo {author}
  {\bibfnamefont {C.}~\bibnamefont {Hille}}, \bibinfo {author} {\bibfnamefont
  {M.}~\bibnamefont {Klett}}, \bibinfo {author} {\bibfnamefont {C.~J.}\
  \bibnamefont {Eckhardt}}, \bibinfo {author} {\bibfnamefont {B.}~\bibnamefont
  {Arzhang}}, \bibinfo {author} {\bibfnamefont {V.}~\bibnamefont {Harkov}},
  \bibinfo {author} {\bibfnamefont {F.-M.}\ \bibnamefont {Le~Régent}},
  \bibinfo {author} {\bibfnamefont {A.}~\bibnamefont {Kirsch}}, \bibinfo
  {author} {\bibfnamefont {Y.}~\bibnamefont {Wang}}, \bibinfo {author}
  {\bibfnamefont {A.~J.}\ \bibnamefont {Kim}}, \bibinfo {author} {\bibfnamefont
  {E.}~\bibnamefont {Kozik}}, \bibinfo {author} {\bibfnamefont {E.~A.}\
  \bibnamefont {Stepanov}}, \bibinfo {author} {\bibfnamefont {A.}~\bibnamefont
  {Kauch}}, \bibinfo {author} {\bibfnamefont {S.}~\bibnamefont {Andergassen}},
  \bibinfo {author} {\bibfnamefont {P.}~\bibnamefont {Hansmann}}, \bibinfo
  {author} {\bibfnamefont {D.}~\bibnamefont {Rohe}}, \bibinfo {author}
  {\bibfnamefont {Y.~M.}\ \bibnamefont {Vilk}}, \bibinfo {author}
  {\bibfnamefont {J.~P.~F.}\ \bibnamefont {LeBlanc}}, \bibinfo {author}
  {\bibfnamefont {S.}~\bibnamefont {Zhang}}, \bibinfo {author} {\bibfnamefont
  {A.-M.~S.}\ \bibnamefont {Tremblay}}, \bibinfo {author} {\bibfnamefont
  {M.}~\bibnamefont {Ferrero}}, \bibinfo {author} {\bibfnamefont
  {O.}~\bibnamefont {Parcollet}},\ and\ \bibinfo {author} {\bibfnamefont
  {A.}~\bibnamefont {Georges}},\ }\bibfield  {title} {\bibinfo {title}
  {Tracking the footprints of spin fluctuations: A multimethod, multimessenger
  study of the two-dimensional hubbard model},\ }\href
  {https://doi.org/10.1103/PhysRevX.11.011058} {\bibfield  {journal} {\bibinfo
  {journal} {Physical Review X}\ }\textbf {\bibinfo {volume} {11}},\ \bibinfo
  {pages} {011058} (\bibinfo {year} {2021})}\BibitemShut {NoStop}%
\bibitem [{\citenamefont {Gauvin-Ndiaye}\ \emph {et~al.}(2023)\citenamefont
  {Gauvin-Ndiaye}, \citenamefont {Lahaie}, \citenamefont {Vilk},\ and\
  \citenamefont {Tremblay}}]{gauvinndiaye2023}%
  \BibitemOpen
  \bibfield  {author} {\bibinfo {author} {\bibfnamefont {C.}~\bibnamefont
  {Gauvin-Ndiaye}}, \bibinfo {author} {\bibfnamefont {C.}~\bibnamefont
  {Lahaie}}, \bibinfo {author} {\bibfnamefont {Y.~M.}\ \bibnamefont {Vilk}},\
  and\ \bibinfo {author} {\bibfnamefont {A.-M.~S.}\ \bibnamefont {Tremblay}},\
  }\bibfield  {title} {\bibinfo {title} {Improved two-particle self-consistent
  approach for the single-band hubbard model in two dimensions},\ }\href
  {https://doi.org/10.1103/PhysRevB.108.075144} {\bibfield  {journal} {\bibinfo
   {journal} {Phys. Rev. B}\ }\textbf {\bibinfo {volume} {108}},\ \bibinfo
  {pages} {075144} (\bibinfo {year} {2023})}\BibitemShut {NoStop}%
\bibitem [{\citenamefont {Gauvin-Ndiaye}\ \emph {et~al.}(2022)\citenamefont
  {Gauvin-Ndiaye}, \citenamefont {Setrakian},\ and\ \citenamefont
  {Tremblay}}]{gauvin2022}%
  \BibitemOpen
  \bibfield  {author} {\bibinfo {author} {\bibfnamefont {C.}~\bibnamefont
  {Gauvin-Ndiaye}}, \bibinfo {author} {\bibfnamefont {M.}~\bibnamefont
  {Setrakian}},\ and\ \bibinfo {author} {\bibfnamefont {A.-M.~S.}\ \bibnamefont
  {Tremblay}},\ }\bibfield  {title} {\bibinfo {title} {Resilient fermi liquid
  and strength of correlations near an antiferromagnetic quantum critical
  point},\ }\href {https://doi.org/10.1103/PhysRevLett.128.087001} {\bibfield
  {journal} {\bibinfo  {journal} {Phys. Rev. Lett.}\ }\textbf {\bibinfo
  {volume} {128}},\ \bibinfo {pages} {087001} (\bibinfo {year}
  {2022})}\BibitemShut {NoStop}%
\bibitem [{\citenamefont {Simard}\ and\ \citenamefont
  {Werner}(2022)}]{simard2022}%
  \BibitemOpen
  \bibfield  {author} {\bibinfo {author} {\bibfnamefont {O.}~\bibnamefont
  {Simard}}\ and\ \bibinfo {author} {\bibfnamefont {P.}~\bibnamefont
  {Werner}},\ }\bibfield  {title} {\bibinfo {title} {Nonequilibrium
  two-particle self-consistent approach},\ }\href
  {https://doi.org/10.1103/PhysRevB.106.L241110} {\bibfield  {journal}
  {\bibinfo  {journal} {Phys. Rev. B}\ }\textbf {\bibinfo {volume} {106}},\
  \bibinfo {pages} {L241110} (\bibinfo {year} {2022})}\BibitemShut {NoStop}%
\bibitem [{\citenamefont {Arya}\ \emph {et~al.}(2015)\citenamefont {Arya},
  \citenamefont {Sriluckshmy}, \citenamefont {Hassan},\ and\ \citenamefont
  {Tremblay}}]{arya2015}%
  \BibitemOpen
  \bibfield  {author} {\bibinfo {author} {\bibfnamefont {S.}~\bibnamefont
  {Arya}}, \bibinfo {author} {\bibfnamefont {P.~V.}\ \bibnamefont
  {Sriluckshmy}}, \bibinfo {author} {\bibfnamefont {S.~R.}\ \bibnamefont
  {Hassan}},\ and\ \bibinfo {author} {\bibfnamefont {A.-M.~S.}\ \bibnamefont
  {Tremblay}},\ }\bibfield  {title} {\bibinfo {title} {Antiferromagnetism in
  the hubbard model on the honeycomb lattice: A two-particle self-consistent
  study},\ }\href {https://doi.org/10.1103/PhysRevB.92.045111} {\bibfield
  {journal} {\bibinfo  {journal} {Phys. Rev. B}\ }\textbf {\bibinfo {volume}
  {92}},\ \bibinfo {pages} {045111} (\bibinfo {year} {2015})}\BibitemShut
  {NoStop}%
\bibitem [{\citenamefont {Aizawa}\ \emph {et~al.}(2015)\citenamefont {Aizawa},
  \citenamefont {Kuroki},\ and\ \citenamefont
  {Yamada}}]{Aizawa_Kuroki_Yamada_2015}%
  \BibitemOpen
  \bibfield  {author} {\bibinfo {author} {\bibfnamefont {H.}~\bibnamefont
  {Aizawa}}, \bibinfo {author} {\bibfnamefont {K.}~\bibnamefont {Kuroki}},\
  and\ \bibinfo {author} {\bibfnamefont {J.-i.}\ \bibnamefont {Yamada}},\
  }\bibfield  {title} {\bibinfo {title} {Enhancement of electron correlation
  due to the molecular dimerization in organic superconductors
  $\ensuremath{\beta}\ensuremath{-}(\mathrm{BDA-TTP}{)}_{2}x\phantom{\rule{0.28em}{0ex}}(x={\mathrm{i}}_{3},
  {\mathrm{sbf}}_{6})$},\ }\href {https://doi.org/10.1103/PhysRevB.92.155108}
  {\bibfield  {journal} {\bibinfo  {journal} {Physical Review B}\ }\textbf
  {\bibinfo {volume} {92}},\ \bibinfo {pages} {155108} (\bibinfo {year}
  {2015})}\BibitemShut {NoStop}%
\bibitem [{\citenamefont {Ogura}\ and\ \citenamefont
  {Kuroki}(2015)}]{Ogura_Kuroki_2015}%
  \BibitemOpen
  \bibfield  {author} {\bibinfo {author} {\bibfnamefont {D.}~\bibnamefont
  {Ogura}}\ and\ \bibinfo {author} {\bibfnamefont {K.}~\bibnamefont {Kuroki}},\
  }\bibfield  {title} {\bibinfo {title} {Asymmetry of superconductivity in
  hole- and electron-doped cuprates: Explanation within two-particle
  self-consistent analysis for the three-band model},\ }\href
  {https://doi.org/10.1103/PhysRevB.92.144511} {\bibfield  {journal} {\bibinfo
  {journal} {Physical Review B}\ }\textbf {\bibinfo {volume} {92}},\ \bibinfo
  {pages} {144511} (\bibinfo {year} {2015})}\BibitemShut {NoStop}%
\bibitem [{\citenamefont {Zantout}\ \emph {et~al.}(2018)\citenamefont
  {Zantout}, \citenamefont {Altmeyer}, \citenamefont {Backes},\ and\
  \citenamefont {Valentí}}]{Zantout_Altmeyer_Backes_Valentí_2018}%
  \BibitemOpen
  \bibfield  {author} {\bibinfo {author} {\bibfnamefont {K.}~\bibnamefont
  {Zantout}}, \bibinfo {author} {\bibfnamefont {M.}~\bibnamefont {Altmeyer}},
  \bibinfo {author} {\bibfnamefont {S.}~\bibnamefont {Backes}},\ and\ \bibinfo
  {author} {\bibfnamefont {R.}~\bibnamefont {Valentí}},\ }\bibfield  {title}
  {\bibinfo {title} {Superconductivity in correlated bedt-ttf molecular
  conductors: Critical temperatures and gap symmetries},\ }\href
  {https://doi.org/10.1103/PhysRevB.97.014530} {\bibfield  {journal} {\bibinfo
  {journal} {Physical Review B}\ }\textbf {\bibinfo {volume} {97}},\ \bibinfo
  {pages} {014530} (\bibinfo {year} {2018})}\BibitemShut {NoStop}%
\bibitem [{\citenamefont {Miyahara}\ \emph {et~al.}(2013)\citenamefont
  {Miyahara}, \citenamefont {Arita},\ and\ \citenamefont
  {Ikeda}}]{miyhara2013}%
  \BibitemOpen
  \bibfield  {author} {\bibinfo {author} {\bibfnamefont {H.}~\bibnamefont
  {Miyahara}}, \bibinfo {author} {\bibfnamefont {R.}~\bibnamefont {Arita}},\
  and\ \bibinfo {author} {\bibfnamefont {H.}~\bibnamefont {Ikeda}},\ }\bibfield
   {title} {\bibinfo {title} {Development of a two-particle self-consistent
  method for multiorbital systems and its application to unconventional
  superconductors},\ }\href {https://doi.org/10.1103/PhysRevB.87.045113}
  {\bibfield  {journal} {\bibinfo  {journal} {Phys. Rev. B}\ }\textbf {\bibinfo
  {volume} {87}},\ \bibinfo {pages} {045113} (\bibinfo {year}
  {2013})}\BibitemShut {NoStop}%
\bibitem [{\citenamefont {Zantout}\ \emph {et~al.}(2019)\citenamefont
  {Zantout}, \citenamefont {Backes},\ and\ \citenamefont
  {Valent\'{\i}}}]{zantout2019}%
  \BibitemOpen
  \bibfield  {author} {\bibinfo {author} {\bibfnamefont {K.}~\bibnamefont
  {Zantout}}, \bibinfo {author} {\bibfnamefont {S.}~\bibnamefont {Backes}},\
  and\ \bibinfo {author} {\bibfnamefont {R.}~\bibnamefont {Valent\'{\i}}},\
  }\bibfield  {title} {\bibinfo {title} {Effect of nonlocal correlations on the
  electronic structure of lifeas},\ }\href
  {https://doi.org/10.1103/PhysRevLett.123.256401} {\bibfield  {journal}
  {\bibinfo  {journal} {Phys. Rev. Lett.}\ }\textbf {\bibinfo {volume} {123}},\
  \bibinfo {pages} {256401} (\bibinfo {year} {2019})}\BibitemShut {NoStop}%
\bibitem [{\citenamefont {Zantout}\ \emph {et~al.}(2021)\citenamefont
  {Zantout}, \citenamefont {Backes},\ and\ \citenamefont
  {Valent{\'\i}}}]{Zantout2021}%
  \BibitemOpen
  \bibfield  {author} {\bibinfo {author} {\bibfnamefont {K.}~\bibnamefont
  {Zantout}}, \bibinfo {author} {\bibfnamefont {S.}~\bibnamefont {Backes}},\
  and\ \bibinfo {author} {\bibfnamefont {R.}~\bibnamefont {Valent{\'\i}}},\
  }\bibfield  {title} {\bibinfo {title} {Two-particle self-consistent method
  for the multi-orbital hubbard model},\ }\href@noop {} {\bibfield  {journal}
  {\bibinfo  {journal} {Annalen der Physik}\ }\textbf {\bibinfo {volume}
  {533}},\ \bibinfo {pages} {2000399} (\bibinfo {year} {2021})}\BibitemShut
  {NoStop}%
\bibitem [{\citenamefont {Davoudi}\ and\ \citenamefont
  {Tremblay}(2006)}]{Davoudi2006}%
  \BibitemOpen
  \bibfield  {author} {\bibinfo {author} {\bibfnamefont {B.}~\bibnamefont
  {Davoudi}}\ and\ \bibinfo {author} {\bibfnamefont {A.-M.~S.}\ \bibnamefont
  {Tremblay}},\ }\bibfield  {title} {\bibinfo {title} {Nearest-neighbor
  repulsion and competing charge and spin order in the extended hubbard
  model},\ }\href {https://doi.org/10.1103/PhysRevB.74.035113} {\bibfield
  {journal} {\bibinfo  {journal} {Phys. Rev. B}\ }\textbf {\bibinfo {volume}
  {74}},\ \bibinfo {pages} {035113} (\bibinfo {year} {2006})}\BibitemShut
  {NoStop}%
\bibitem [{\citenamefont {Davoudi}\ and\ \citenamefont
  {Tremblay}(2007)}]{Davoudi2007}%
  \BibitemOpen
  \bibfield  {author} {\bibinfo {author} {\bibfnamefont {B.}~\bibnamefont
  {Davoudi}}\ and\ \bibinfo {author} {\bibfnamefont {A.-M.~S.}\ \bibnamefont
  {Tremblay}},\ }\bibfield  {title} {\bibinfo {title} {Non-perturbative
  treatment of charge and spin fluctuations in the two-dimensional extended
  hubbard model: Extended two-particle self-consistent approach},\ }\href
  {https://doi.org/10.1103/PhysRevB.76.085115} {\bibfield  {journal} {\bibinfo
  {journal} {Phys. Rev. B}\ }\textbf {\bibinfo {volume} {76}},\ \bibinfo
  {pages} {085115} (\bibinfo {year} {2007})}\BibitemShut {NoStop}%
\bibitem [{\citenamefont {Davoudi}\ \emph {et~al.}(2008)\citenamefont
  {Davoudi}, \citenamefont {Hassan},\ and\ \citenamefont
  {Tremblay}}]{Davoudi2008}%
  \BibitemOpen
  \bibfield  {author} {\bibinfo {author} {\bibfnamefont {B.}~\bibnamefont
  {Davoudi}}, \bibinfo {author} {\bibfnamefont {S.~R.}\ \bibnamefont
  {Hassan}},\ and\ \bibinfo {author} {\bibfnamefont {A.-M.~S.}\ \bibnamefont
  {Tremblay}},\ }\bibfield  {title} {\bibinfo {title} {Competition between
  charge and spin order in the $t\text{\ensuremath{-}}u\text{\ensuremath{-}}v$
  extended hubbard model on the triangular lattice},\ }\href
  {https://doi.org/10.1103/PhysRevB.77.214408} {\bibfield  {journal} {\bibinfo
  {journal} {Phys. Rev. B}\ }\textbf {\bibinfo {volume} {77}},\ \bibinfo
  {pages} {214408} (\bibinfo {year} {2008})}\BibitemShut {NoStop}%
\bibitem [{\citenamefont {Shinaoka}\ \emph {et~al.}(2017)\citenamefont
  {Shinaoka}, \citenamefont {Otsuki}, \citenamefont {Ohzeki},\ and\
  \citenamefont {Yoshimi}}]{Shinaoka2017}%
  \BibitemOpen
  \bibfield  {author} {\bibinfo {author} {\bibfnamefont {H.}~\bibnamefont
  {Shinaoka}}, \bibinfo {author} {\bibfnamefont {J.}~\bibnamefont {Otsuki}},
  \bibinfo {author} {\bibfnamefont {M.}~\bibnamefont {Ohzeki}},\ and\ \bibinfo
  {author} {\bibfnamefont {K.}~\bibnamefont {Yoshimi}},\ }\bibfield  {title}
  {\bibinfo {title} {Compressing green's function using intermediate
  representation between imaginary-time and real-frequency domains},\ }\href
  {https://doi.org/10.1103/PhysRevB.96.035147} {\bibfield  {journal} {\bibinfo
  {journal} {Phys. Rev. B}\ }\textbf {\bibinfo {volume} {96}},\ \bibinfo
  {pages} {035147} (\bibinfo {year} {2017})}\BibitemShut {NoStop}%
\bibitem [{\citenamefont {Li}\ \emph {et~al.}(2020)\citenamefont {Li},
  \citenamefont {Wallerberger}, \citenamefont {Chikano}, \citenamefont {Yeh},
  \citenamefont {Gull},\ and\ \citenamefont {Shinaoka}}]{Li2020}%
  \BibitemOpen
  \bibfield  {author} {\bibinfo {author} {\bibfnamefont {J.}~\bibnamefont
  {Li}}, \bibinfo {author} {\bibfnamefont {M.}~\bibnamefont {Wallerberger}},
  \bibinfo {author} {\bibfnamefont {N.}~\bibnamefont {Chikano}}, \bibinfo
  {author} {\bibfnamefont {C.-N.}\ \bibnamefont {Yeh}}, \bibinfo {author}
  {\bibfnamefont {E.}~\bibnamefont {Gull}},\ and\ \bibinfo {author}
  {\bibfnamefont {H.}~\bibnamefont {Shinaoka}},\ }\bibfield  {title} {\bibinfo
  {title} {Sparse sampling approach to efficient ab initio calculations at
  finite temperature},\ }\href {https://doi.org/10.1103/PhysRevB.101.035144}
  {\bibfield  {journal} {\bibinfo  {journal} {Phys. Rev. B}\ }\textbf {\bibinfo
  {volume} {101}},\ \bibinfo {pages} {035144} (\bibinfo {year}
  {2020})}\BibitemShut {NoStop}%
\bibitem [{\citenamefont {Wallerberger}\ \emph {et~al.}(2023)\citenamefont
  {Wallerberger}, \citenamefont {Badr}, \citenamefont {Hoshino}, \citenamefont
  {Huber}, \citenamefont {Kakizawa}, \citenamefont {Koretsune}, \citenamefont
  {Nagai}, \citenamefont {Nogaki}, \citenamefont {Nomoto}, \citenamefont
  {Mori}, \citenamefont {Otsuki}, \citenamefont {Ozaki}, \citenamefont
  {Plaikner}, \citenamefont {Sakurai}, \citenamefont {Vogel}, \citenamefont
  {Witt}, \citenamefont {Yoshimi},\ and\ \citenamefont
  {Shinaoka}}]{Wallerberger2023}%
  \BibitemOpen
  \bibfield  {author} {\bibinfo {author} {\bibfnamefont {M.}~\bibnamefont
  {Wallerberger}}, \bibinfo {author} {\bibfnamefont {S.}~\bibnamefont {Badr}},
  \bibinfo {author} {\bibfnamefont {S.}~\bibnamefont {Hoshino}}, \bibinfo
  {author} {\bibfnamefont {S.}~\bibnamefont {Huber}}, \bibinfo {author}
  {\bibfnamefont {F.}~\bibnamefont {Kakizawa}}, \bibinfo {author}
  {\bibfnamefont {T.}~\bibnamefont {Koretsune}}, \bibinfo {author}
  {\bibfnamefont {Y.}~\bibnamefont {Nagai}}, \bibinfo {author} {\bibfnamefont
  {K.}~\bibnamefont {Nogaki}}, \bibinfo {author} {\bibfnamefont
  {T.}~\bibnamefont {Nomoto}}, \bibinfo {author} {\bibfnamefont
  {H.}~\bibnamefont {Mori}}, \bibinfo {author} {\bibfnamefont {J.}~\bibnamefont
  {Otsuki}}, \bibinfo {author} {\bibfnamefont {S.}~\bibnamefont {Ozaki}},
  \bibinfo {author} {\bibfnamefont {T.}~\bibnamefont {Plaikner}}, \bibinfo
  {author} {\bibfnamefont {R.}~\bibnamefont {Sakurai}}, \bibinfo {author}
  {\bibfnamefont {C.}~\bibnamefont {Vogel}}, \bibinfo {author} {\bibfnamefont
  {N.}~\bibnamefont {Witt}}, \bibinfo {author} {\bibfnamefont {K.}~\bibnamefont
  {Yoshimi}},\ and\ \bibinfo {author} {\bibfnamefont {H.}~\bibnamefont
  {Shinaoka}},\ }\bibfield  {title} {\bibinfo {title} {sparse-ir: Optimal
  compression and sparse sampling of many-body propagators},\ }\href
  {https://doi.org/https://doi.org/10.1016/j.softx.2022.101266} {\bibfield
  {journal} {\bibinfo  {journal} {SoftwareX}\ }\textbf {\bibinfo {volume}
  {21}},\ \bibinfo {pages} {101266} (\bibinfo {year} {2023})}\BibitemShut
  {NoStop}%
\bibitem [{\citenamefont {Lessnich}(2023)}]{Lessnich_TPSC_2023}%
  \BibitemOpen
  \bibfield  {author} {\bibinfo {author} {\bibfnamefont {D.}~\bibnamefont
  {Lessnich}},\ }\href {https://github.com/Dominik-Lessnich/TPSC_KMH} {\bibinfo
  {title} {{TPSC for the Kane-Mele-Hubbard model}}} (\bibinfo {year}
  {2023})\BibitemShut {NoStop}%
\bibitem [{\citenamefont {Nourafkan}\ and\ \citenamefont
  {Tremblay}(2018)}]{noufrakan2018}%
  \BibitemOpen
  \bibfield  {author} {\bibinfo {author} {\bibfnamefont {R.}~\bibnamefont
  {Nourafkan}}\ and\ \bibinfo {author} {\bibfnamefont {A.-M.~S.}\ \bibnamefont
  {Tremblay}},\ }\bibfield  {title} {\bibinfo {title} {Hall and faraday effects
  in interacting multiband systems with arbitrary band topology and spin-orbit
  coupling},\ }\href {https://doi.org/10.1103/PhysRevB.98.165130} {\bibfield
  {journal} {\bibinfo  {journal} {Phys. Rev. B}\ }\textbf {\bibinfo {volume}
  {98}},\ \bibinfo {pages} {165130} (\bibinfo {year} {2018})}\BibitemShut
  {NoStop}%
\bibitem [{\citenamefont {Ishikawa}\ and\ \citenamefont
  {Matsuyama}(1987)}]{Ishikawa1987}%
  \BibitemOpen
  \bibfield  {author} {\bibinfo {author} {\bibfnamefont {K.}~\bibnamefont
  {Ishikawa}}\ and\ \bibinfo {author} {\bibfnamefont {T.}~\bibnamefont
  {Matsuyama}},\ }\bibfield  {title} {\bibinfo {title} {A microscopic theory of
  the quantum hall effect},\ }\href
  {https://doi.org/https://doi.org/10.1016/0550-3213(87)90160-X} {\bibfield
  {journal} {\bibinfo  {journal} {Nuclear Physics B}\ }\textbf {\bibinfo
  {volume} {280}},\ \bibinfo {pages} {523} (\bibinfo {year}
  {1987})}\BibitemShut {NoStop}%
\bibitem [{\citenamefont {Bergeron}\ \emph {et~al.}(2011)\citenamefont
  {Bergeron}, \citenamefont {Hankevych}, \citenamefont {Kyung},\ and\
  \citenamefont {Tremblay}}]{bergeron2011}%
  \BibitemOpen
  \bibfield  {author} {\bibinfo {author} {\bibfnamefont {D.}~\bibnamefont
  {Bergeron}}, \bibinfo {author} {\bibfnamefont {V.}~\bibnamefont {Hankevych}},
  \bibinfo {author} {\bibfnamefont {B.}~\bibnamefont {Kyung}},\ and\ \bibinfo
  {author} {\bibfnamefont {A.-M.~S.}\ \bibnamefont {Tremblay}},\ }\bibfield
  {title} {\bibinfo {title} {Optical and dc conductivity of the two-dimensional
  hubbard model in the pseudogap regime and across the antiferromagnetic
  quantum critical point including vertex corrections},\ }\href
  {https://doi.org/10.1103/PhysRevB.84.085128} {\bibfield  {journal} {\bibinfo
  {journal} {Phys. Rev. B}\ }\textbf {\bibinfo {volume} {84}},\ \bibinfo
  {pages} {085128} (\bibinfo {year} {2011})}\BibitemShut {NoStop}%
\bibitem [{\citenamefont {Markov}\ \emph {et~al.}(2019)\citenamefont {Markov},
  \citenamefont {Rohringer},\ and\ \citenamefont {Rubtsov}}]{Markov2019}%
  \BibitemOpen
  \bibfield  {author} {\bibinfo {author} {\bibfnamefont {A.~A.}\ \bibnamefont
  {Markov}}, \bibinfo {author} {\bibfnamefont {G.}~\bibnamefont {Rohringer}},\
  and\ \bibinfo {author} {\bibfnamefont {A.~N.}\ \bibnamefont {Rubtsov}},\
  }\bibfield  {title} {\bibinfo {title} {Robustness of the topological
  quantization of the hall conductivity for correlated lattice electrons at
  finite temperatures},\ }\href {https://doi.org/10.1103/PhysRevB.100.115102}
  {\bibfield  {journal} {\bibinfo  {journal} {Phys. Rev. B}\ }\textbf {\bibinfo
  {volume} {100}},\ \bibinfo {pages} {115102} (\bibinfo {year}
  {2019})}\BibitemShut {NoStop}%
\end{thebibliography}%

\end{document}